\documentclass[12 pt, onehalfspacing]{amsart}
\usepackage{amsthm,amsmath}
\usepackage{amsfonts}
\usepackage{amsmath}
\usepackage{fancyhdr}
\usepackage[T1]{fontenc}
\usepackage[latin9]{inputenc}
\usepackage{mathabx}
\usepackage{color}
\usepackage{float}
\usepackage{mathrsfs}
\usepackage{layout}
\usepackage{slashed}
\usepackage{amsthm}
\usepackage{mathptmx}
\usepackage{amstext}
\usepackage{amssymb}
\usepackage{graphicx}
\usepackage{stmaryrd}
\usepackage{graphicx}
\usepackage{setspace}
\usepackage[export]{adjustbox}
\usepackage{esint}
\usepackage[strict]{changepage}
\RequirePackage[colorlinks=true,citecolor=blue,urlcolor=blue]{hyperref}
\usepackage[authoryear]{natbib}
\usepackage{ulem}
\usepackage{enumitem}
\usepackage{ifpdf}
\usepackage{pdflscape}
\usepackage[figure]{hypcap}
\usepackage{fancyhdr}
\usepackage{bbm}
\usepackage{geometry}
\usepackage[parfill]{parskip}
\allowdisplaybreaks
\theoremstyle{plain}
\newtheorem{thm}{\protect\theoremname}
\theoremstyle{definition}

\theoremstyle{plain}
\newtheorem{prop}[thm]{Proposition}
\theoremstyle{definition}

\theoremstyle{plain}
\newtheorem{defn*}{}
\theoremstyle{plain}

\theoremstyle{plain}

\theoremstyle{plain}

\ifpdf 
 \IfFileExists{lmodern.sty}{\usepackage{lmodern}}{}
\fi 
\let\myTOC\tableofcontents
\renewcommand\tableofcontents{  \frontmatter
  \pdfbookmark[1]{\contentsname}{}
  \myTOC
  \mainmatter }

\usepackage{cleveref}

\begin{document}

		\title{Heterogeneity and the Dynamic Effects of Aggregate Shocks\\ }
		\author{Andreas Tryphonides\\ University of Cyprus}

		\address{P.O. Box 20537, Nicosia, Cyprus. }

		\email{tryfonidis.antreas@ucy.ac.cy}

		\date{Mar 2020, Previous versions: Feb 2019, Nov 2019}
\thanks{\tiny{Parts of this paper (Section 3) draw from chapter 2 of my PhD thesis (EUI, Sep 2016), and the paper circulated under the title "Surveys and Margins of Adjustment in Heterogeneous Agent Economies".  I thank Fabio Canova, Peter Reinhard Hansen, Frank Schorfheide and Giuseppe Ragusa for comments and suggestions on the thesis chapter and the participants at the Society for Economic Dynamics conference (2019, St Louis) and the International Association for Applied Econometrics (2019, Nicosia) for comments. Any errors are my own.}}

	\setcounter{page}{1}
	\pagenumbering{arabic}
	\pagestyle{plain}
	\thispagestyle{empty}

\begin{abstract}  Using a semi-structural approach, the paper identifies how heterogeneity and financial frictions affect the transmission of aggregate shocks. Approximating a heterogeneous agent model around the representative agent allocation can successfully trace the aggregate and distributional dynamics and can be consistent with alternative mechanisms. Employing Spanish macroeconomic data as well as firm and household survey data, the paper finds  that frictions on both consumption and  investment have rich interactions with aggregate shocks. The response of heterogeneity amplifies or dampens these effects depending on the type of the shock. Both dispersion in consumption shares and the marginal revenue product of firms, as well as the proportion of investment constrained firms are key determinants of the fiscal multiplier.
	\bigskip \newline
{\small 	Keywords:  Heterogeneity, Aggregate Shocks, Micro Data. \newline JEL Classification: C5, C6, D52} 
\end{abstract}
 \clearpage

	\maketitle

\newpage
\normalsize


\setcounter{page}{1}

\section{Introduction}

During the last three decades, the literature on heterogeneous agent  models has explored various types of frictions and heterogeneity to explain stylized facts obtained from firm and household level data. While recent modeling advances have made it possible to evaluate the business cycle implications of agent heterogeneity\footnote{See i.e. \citet{RePEc:nbr:nberch:13927, doi:10.3982/QE714}.}, identifying the interaction between financial frictions, aggregate shocks and heterogeneity from the data is still a challenging task. Credibly producing such evidence can shed light on what kind of business cycle facts we should strive to explain or heterogeneous agents models should be able to replicate. Along this line of research, a recent strand of the literature focuses on estimates of the distribution of the marginal propensities to consume, which can be used as a sufficient statistic for computing the impact effect of a policy shock in economies with idiosyncratic risk\footnote{See \citet{doi:10.3982/ECTA11883,RePEc:nbr:nberwo:25020,NBERw23451} for a partial and a general equilibrium setting and \citet{doi:10.1146/annurev.economics.050708.142910} for a review of the sufficient statistics approach to program evaluation.}. Nevertheless, aggregate shocks can also alter the way heterogeneity matters for aggregate fluctuations.

This paper develops a methodology to identify the role of financial frictions and heterogeneity in the determination of the economy's response to aggregate shocks.  Using an approximated heterogeneous agent (HA) model around the representative agent (RA) allocation, it illustrates that financial frictions generate distortions to aggregates along different margins. The latter depend on cross sectional moments e.g. the cross sectional dispersion in consumption and the marginal revenue product of firms, as well as the extensive and intensive margin of adjustment of households and firms which are financially constrained e.g. the proportion that is indeed constrained and by how much, on average. 
These moments are either directly observable but possibly at a lower frequency than macroeconomic variables, or they are partially unobserved, but can be uncovered using information from aggregated qualitative and quantitative survey data. Qualitative data include questions that directly reveal the status of each respondent while quantitative micro data provide measures of cross sectional dispersion over time.

The paper utilizes restrictions that hold across alternative models, hence the parameters of the benchmark representative agent model and the wedges that are generated by heterogeneity and frictions are not necessarily point identified. Nevertheless, the approximation used imposes additional cross equation restrictions, and the additional information contained in survey data permits  the identification of the interaction of heterogeneity and the intensive and extensive margins of adjustment with aggregate shocks. Furthermore, the link between measurements and all unobservables is formulated in a Gaussian state space model, which explicitly controls for both the fact that surveys are available at different frequencies and for the presence of errors in measurements.  

The methodology is used to identify distortions due to consumer and firm financial frictions in the Spanish economy. Empirical results indicate rich interactions. Shocks to the ex-ante real interest rate have significant effects on these margins while government spending shocks seem to explain more than a third of the unconditional variation in most of the observables. The response of output to spending shocks is significantly affected by heterogeneity, either in consumer spending or the marginal revenue product of firms while the extensive margin of investment constrained firms has the most important contribution in terms of size. There are also important feedback effects between heterogeneity and aggregate shocks, as well as the proportion of constrained consumers and firms. 

Another finding is that despite that the model does not hard-wire a New Keynesian mechanism, markups are found to be countercyclical conditioning on government spending shocks, while there is no clear evidence on the cyclicality with respect to the real interest rate and aggregate productivity.

\subsection{Contribution and contacts with the literature}
This paper relates to the literature that compares the predictions of HANK (Heterogeneous Agent New Keynesian) models and approximations to them like the TANK (Two agent NK) model and their differences to the representative agent model (RANK). \cite{pub.1105955662} illustrate that the extent of the difference in predictions depends on the type of shock we consider. \cite{GaliDeb} use a second order approximation and find that the  TANK model can quantitatively approximate the dynamics of the HANK model reasonably well. Whether variations in consumption heterogeneity and the corresponding proportion of constrained consumers matter empirically, quoting \cite{GaliDeb},  "remains an open question". 

This paper employs a similar approximation and offers a general methodology to tackle this issue while it is not tied to a specific mechanism. More particularly, measuring the proportion of agents that are constrained over time provides evidence on the variability of this proportion and how it interacts with aggregate shocks. In TANK models, this proportion is taken as fixed and exogenous. Second, as shown in Section 3, the representation used captures aggregate precautionary effects from anticipating binding constraints in the future, something that is also not taken into account by the TANK model. {Accounting for these effects can shed light on their relative importance\footnote{Precautionary savings and binding constraints are expected to generate similar behavior, see e.g. \citet{10.1257/jep.15.3.23,NBERw8496}.}.} Moreover, this proportion can also provide information for identifying the variance of consumption when it is latent at higher frequencies as well as when the variance is primarily affected by the constrained households.

Methodology-wise, the paper asks two interrelated questions: (a) Given the approximation, does there exist a corresponding law of motion within which the complete markets allocation is smoothly embedded? (b) Is this approximation adequate for \textit{empirically} approximating the law of motion of a HA economy with aggregate uncertainty? Providing an answer to the first question is important for a few reasons: If the complete markets' allocation is a smooth special case of the aggregate law of motion of the approximate HA model then a likelihood function can be easily constructed. Moreover, if the additional components generated by heterogeneity are tightly connected to the structural parameters, they can be linked to any kind of data that can be potentially informative, which matters for identification. For the second question, showing that this is a good approximation for the workhorse model is re-assuring and a useful first step. 

The answer is affirmative for both questions. The paper shows that there exists a law of motion that is a distortion to the representative agent allocation which is linearly related to the distortion of the aggregate first order conditions. Moreover, the mapping between the former and the latter indeed imposes additional cross equation restrictions. 
Furthermore,  within the context of the benchmark \citet{krusellsmith} economy (KS hereafter), the empirical approximation is precise, under parameterizations that imply alternative degrees of self-insurance. For the benchmark parameterization, where risk aversion and the discount factor are reasonably "high" and on par with standard values used in the literature for model calibration, consumption variance  is mainly driven by aggregate shocks. When both are low enough,   consumption variance cannot  be entirely explained by aggregate shocks, yet all macroeconomic variables are still well approximated. What is more important is that the paper's approach controls for distributional effects should approximate aggregation does not hold and for their dependence on aggregate shocks.

The paper's approximation to the HA model generates cross sectional moments as well as aggregate wedges in investment and consumption. One important cross sectional moment that arises is the dispersion in consumption shares, an observation that goes back to \citet{RePEc:ucp:jpolec:v:104:y:1996:i:2:p:219-40}. Another moment, more specific to the model used in estimation, is the cross sectional dispersion in the marginal product of firms, which is the outcome of heterogeneous capital utilization rates. In turn, consumption and investment are distorted from those agents (consumers and/or  firms) that are hitting their constraint, which can be decomposed into an intensive and an extensive margin without imposing any additional assumption. The additional step is to link these wedges to aggregate shocks which provides direct evidence on their relationship but also the quantitative impact on aggregate outcomes. As illustrated in the application, this is relevant for quantifying the effects of fiscal policy during recessions on aggregate consumption. 
This contributes to the voluminous literature on the effects of government purchases to macroeconomic aggregates, summarized in \citet{10.1257/jep.33.2.89}.

In recent work, \citet{RePEc:red:sed018:588} measure the implications of incomplete markets for the business cycle by utilizing that trading constraints lead to a representative agent aggregate Euler equation with a time varying discount factor\footnote{This is pinned down by the unconstrained agent with the highest bond valuation. The authors also refer to similar results of \citet{wernig2015} and \citet{RePEc:eee:jetheo:v:145:y:2010:i:1:p:1-41}.}. While I also consider a distorted Euler equation, the employed representation is linked to both micro and data across the business cycle in one step. This informs parameter estimates as well as identifying the different margins and their dependence on aggregate shocks. 

\cite{Chang2018HeterogeneityAA} propose a  Bayesian reduced form methodology for identifying the interaction of  heterogeneity and macroeconomic outcomes. \citet{LiuMoller} propose a Bayesian full information method to estimate a HA model using macro and micro data. This paper is similar in spirit and complements these papers. It differs by employing the second order approximation to a HA model, which easily lends itself to conventional likelihood based methods. Moreover, being quasi-structural, it does not adhere to a specific mechanism that generates distortions and heterogeneity and can therefore identify interactions that could otherwise be ignored due to  misspecification, while it allows for computing certain counterfactuals.    

In a companion paper, \citet{Tryphonides2020} studies credible identification of structural parameters  through moment inequalities (limited information) and the role of the extensive margin in improving inference by shrinking the set of admissible estimates. This paper builds on this approach by developing the "full information" counterpart\footnote{We cannot call this a proper full information approach as agents are boundedly rational, yet the econometrician and the agents have the same information set with an infinite sample.}, which provides a general framework that can be used to do likelihood based inference in this setting. This is important as state space methods allow for greater flexibility in linking aggregated micro-surveys to different components of the approximated HA model.

 The rest of the paper is structured as follows: Section 2 summarizes the KS model as well as the second order approximation to illustrate the issue at hand. Section 3 presents the result on the link between aggregate wedges and the equilibrium law of motion. Section 4 presents the experiments on the quality of the empirical approximation. Section 5 presents the extensive study of the role of heterogeneity and financial frictions in Spain and Section 6 concludes. 
%
\section{The \citet{krusellsmith} economy} 
In order to build intuition for the approximation of the HA economy, I first introduce the KS model.
Adopting the formulation in \citet{MALIAR201042, denhaanjudd}, the KS economy  features a standard capital accumulation problem under incomplete markets, where the first order conditions for each household are as follows: 
\begin{eqnarray*}
	c^{-\omega}_{i,t}&=& \beta\mathbb{E}_{t}(1-\delta + r^{k}_{t+1})c^{-\omega}_{i,t+1}+\lambda_{i,t}\\
	k_{i,t+1} &=& (1-\tau)w_{t}\bar{l}\epsilon_{i} + \nu_{t} w_{t} (1-\epsilon_{i})+(1-\delta+r^{k}_{t})k_{i,t}-c_{i,t}\\
	k_{i,t+1}&\geq& 0 
\end{eqnarray*} 
where $\epsilon_{i}\in\{0,1\}$ is the idiosyncratic employment shock, $\bar{l}$ the time endowment and $v_{t}$ the fraction of the wage that is received as unemployment benefit. 
The first equation is the standard Euler equation which is distorted due to the occasionally binding borrowing constraint, while the second equation is the budget constraint, which nests the case of employment and unemployment.

Perfectly competitive markets imply that the interest and wage rates are equal to the marginal products of capital and labor: \[r^{k}_{t}=Z_{t}\alpha\left(\frac{K_{t}}{\bar{l}L_{t}}\right)^{\alpha-1},  \quad w_{t}=Z_{t}(1-\alpha)\left(\frac{K_{t}}{\bar{l}L_{t}}\right)^{\alpha}\]
while the labor income tax rate is designed to be equal to $\tau = \frac{\nu_{t} u_{t}}{\bar{l}L_{t}}$. 

The only source of aggregate risk is the aggregate productivity shock $Z_{t}$ which takes two values, $\left\{Z_{low},Z_{high}\right\}$. In turn, the unemployment rate $u_{t}$ depends on the productivity shock as well, and takes two possible values, $\left\{u_{low},u_{high}\right\}$, with transition matrix $P=(p_{ll},p_{lh};p_{hl},p_{hh})$. Finally, full employment is normalized to one, thus $L = 1-u$.

\subsubsection*{Second Order Approximation}
Under complete markets, individual consumption is proportional to aggregate consumption and the consumption shares do not vary over time. A reasonable point of approximation to individual marginal utility would therefore be the marginal utility of the representative agent\footnote{See also \citet{GaliDeb}, \citet{Tryphonides2020}.}: \vspace{-0.1in}\[c^{-\omega}_{i,t} \approx  C^{-\omega}_{t} -\omega C^{-\omega-1}_{t}(c_{i,t}-C_{t}) + \frac{\omega(\omega+1)}{2}C^{-\omega-2}_{t}(c_{i,t}-C_{t})^{2}\]
Denote by $s_{i,t}=(\epsilon_{i,t},k_{i,t})$ the vector of idiosyncratic states and the distribution of idiosyncratic states conditional on aggregate states by $p(s_{i,t}|S_{t})$. In a rational expectations equilibrium, agent expectations are formed using $p(s_{i,t+1},S_{t+1}|s_{i,t},S_{t})$. Aggregating the Euler equation using  $p(s_{i,t}|S_{t})$ yields the following condition:\small \begin{eqnarray}
C_{t}^{-\omega}& =& \beta\mathbb{E}_{t}(1-\delta+ r^{k}_{t+1})\left(\frac{\Xi_{t+1}}{\Xi_{t}}\right)C^{-\omega}_{t+1}+\int \lambda_{i,t}p(s_{i,t}|S_{t})ds_{i,t}\nonumber \\
& :=& \beta\mathbb{E}_{t}(1-\delta+ r^{k}_{t+1})\left(\frac{\Xi_{t+1}}{\Xi_{t}}\right)C^{-\omega}_{t+1}+\mu_{c,t}\label{eq:AppEul}
\end{eqnarray} \normalsize
where\footnote{Under complete markets, marginal utility growth is equal across agents and the RA approximation becomes exact, $	C^{-\omega}_{t}=\beta\mathbb{E}_{t}C^{-\omega}_{t+1}(1-\delta + r^{k}_{t+1})$.}\vspace{-0.1 in}\small
\[C_{t+1}\equiv \int c_{i,t+1 }p(s_{i,t+1},s_{i,t}|S_{t+1},S_{t})d(s_{i,t+1},s_{i,t}), \quad  C_{t}\equiv \int c_{i,t }p(s_{i,t}|S_{t})ds_{i,t}\] \[\Xi_{t} \equiv 1+\frac{\omega(\omega+1)}{2}C^{-2}_{t}Var(c_{i,t}), \quad\Xi_{t,t+1} \equiv 1+ \frac{\omega(\omega+1)}{2}C^{-2}_{t+1}Var(c_{i,t+1})\]
\[Var(c_{i,t}) =  \int (c_{i,t}-C_{t})^2p(s_{i,t}|S_{t})d s_{i,t}\]\[Var(c_{i,t+1}) =  \int (c_{i,t+1}-C_{t+1})^2p(s_{i,t+1},s_{i,t}|S_{t+1},S_{t})d (s_{i,t+1},s_{i,t})\]

\normalsize

The aggregate distortion to \eqref{eq:AppEul} may not vanish in general equilibrium, and will depend on the aggregate states, $S_{t}$. Moreover, notice that $\mu^{c}_{t}$ involves adjustments on the extensive and the intensive margin i.e. it depends on the proportion of households that adjust $(i:\lambda_{i,t}>0)$ and by how much. 

The aggregated Euler equation can now be analysed, both non-linearly or using log-linearization techniques. We will proceed with the latter. More particularly, the  log-linearized version of \eqref{eq:AppEul} is as follows:
\begin{eqnarray}\tilde{C}_{t} &=& \beta(1-\delta+r^{k}_{ss})\mathbb{E}_{t} \left(\tilde{C}_{t+1}-\frac{1}{\omega}\Delta\tilde{\Xi}_{t+1}-\frac{r^{k}_{ss}}{\omega(1-\delta+r^{k}_{ss})}\tilde{r}^{k}_{t}\right)+\tilde{\mu}_{c,t}\label{eq:AppEul2}\end{eqnarray}
and together with the rest of the conditions they form a rational expectations system, as in the RA case.
Note that as is evident form the second order approximation, the cross sectional moments can be part of the state space e.g. in the second order approximation, the cross sectional variance appears as an additional state variable. In the case in which the whole distribution becomes a state variable, then an  uncountable number of moments could be in principle used to represent this economy, and hence the computational issue that \citet{krusellsmith} have tried to tackle. 

Before presenting the complete log-linearized system of the KS economy, I present the theoretical results about the relation between the law of motion of the representative agent and the approximated HA model. These results are general, in the sense that they do not only apply to the HA model, but to any model that can be described as a distortion to some benchmark. In this case the benchmark case in the RA allocation\footnote{This is also related to the wedge literature initiated by \citet{ECTA:ECTA768} (CKM) who identify wedges
	in the optimality conditions of a frictionless model that produce the same
	equilibrium allocations in economies with specific parametric choices for
	the frictions. We also take a frictionless model as a benchmark but contrary
	to CKM we show that there is a precise mapping from wedges to distortions in the decision rules, and thus additional cross equation restrictions that are important for identifying how the distortions depend on aggregate shocks.}.
\section{The distorted equilibrium law of motion}
To generalize the analysis I adopt the following notation: Let $X_{t}$ denote the aggregate endogenous variables and $Z_{t}$ the aggregate shocks e.g. $S_{t}:=(X_{t},Z_{t})'$. Then the system of aggregate equilibrium conditions becomes as follows:
\begin{eqnarray}
G(\theta)X_{t}&=&F(\theta)\mathbb{E}_{t}(X_{t+1}|X_{t})+L(\theta)Z_{t}+{\tilde{\mu} _{t}}\label{eq:exp_sys_pert}\\
Z_{t} &=&R(\theta )Z_{t-1}+\epsilon _{t}  \nonumber
\end{eqnarray} 
Under complete markets, the model reduces to the frictionless representative agent model, where $\tilde{\mu}_{t}=0$ and the cross sectional moments in $X_{t}$ become redundant. I thus assume that a subset of parameters in $\theta$, say $\theta_{2}$, is also set to zero. In the context of the KS model, this can correspond to eliminating $\tilde{\Xi}_{t}$, the log-linearized version of the cross sectional consumption variance, which does not appear under complete markets.
The vector  $\theta $ is thus partitioned into two subsets, $(\theta _{1},\theta
_{2})$, and the frictionless economy is
summarized by:
\begin{eqnarray}  \label{eq:exp_sys}
{G}(\theta _{1},0)X_{t} &=&F(\theta _{1},0)\mathbb{E}%
_{t}(X_{t+1}|X_{t},Z_{t})+L(\theta _{1},0)Z_{t}  \label{exp_eq_no_frict}
\\
Z_{t} &=&R(\theta )Z_{t-1}+\epsilon _{t} \nonumber
\end{eqnarray}
 Therefore, $X^{\star}_{t}=\underset{n_{x}\times n_{x}}{P^{\ast }(\theta
 	_{1},0)}X_{t-1}+\underset{n_{x}\times n_{z}}{Q^{\ast }(\theta _{1},0)}Z_{t}$ is a Rational Expectations equilibrium if there exist unique matrices ${P^{\ast }(\theta
			_{1},0)},{Q^{\ast }(\theta _{1},0)}$ such that the following conditions are satisfied\footnote{See
			for example \citet{Marimon_Comp}.}:  \vspace{-0.4 in}\begin{spacing}{1.4}
		\begin{eqnarray*}
			(F(\theta_{1},0)P^{\ast }(\theta _{1},0)-G^{\ast }(\theta _{1},0))P^{\ast
			}(\theta _{1},0)=0 &&\\
			({R(\theta )^{T}}\otimes F(\theta _{1},0)+I_{z}\otimes (F(\theta
			_{1},0)P^{\ast }(\theta _{1},0)-G^{\ast }(\theta _{1},0)))vec(Q(\theta
			_{1},0))&&\\
			=-vec(L(\theta _{1},0))&&
		\end{eqnarray*}\end{spacing} 			  
Correspondingly, we can always rewrite \eqref{eq:exp_sys_pert} as follows:  
\begin{eqnarray}
	G(\theta_1,0)X_{t}&=&F(\theta_1,0)\mathbb{E}_{t}(X_{t+1}|X_{t})+L(\theta_1,0)Z_{t}+{\mu _{t}} \label{eq:Exp_Sys_Dist}
\end{eqnarray}
where $\mu_{t}$ subsumes both $\tilde{\mu}_{t}$ and the difference between \eqref{eq:exp_sys_pert} and  \eqref{exp_eq_no_frict} due to setting $\theta_{2}=0$. We could have differentiated between distortions that appear in the expectational component (e.g. $\Xi_{t}$ in the KS economy) and the rest of the model. Nevertheless, while this complicates notation,  it makes no difference to the main result that follows as in equilibrium, $\Xi_{t}$ will be a function of the aggregate states. The next proposition characterizes the relationship between ${\mu} _{t}$
and a set of candidate aggregate decision rules that are distortions of the RA law of motion. \vspace{-0.3 in}
\begin{spacing}{1.4}
		\begin{prop} For matrices $(M_{1},M_{2})$ such that ${\mu}_{t} = M_{1}X_{t-1} + M_{2}Z_{t}$: 
		\begin{enumerate} 
			\item There exists a distorted aggregate rule $X_{t}=X_{t}^{\star}+{\lambda}_{t}$ which satisfies \eqref{eq:Exp_Sys_Dist}
			\item There exists a matrix $H$ such that ${\lambda}_{t} = H{\mu}_{t}$, where the following restrictions hold: \vspace{-0.1 in}\small
			\begin{eqnarray*} vec(H) &=& \left(M_{1}'\otimes G(\theta_{1},0) - (M_{1}P^{\star}(\theta_{1},\theta_{2}))'\otimes F(\theta_{1},0))\right)^{-1}vec(M_{1})\\
					vec(M_{2}) &=& \left(I_{n_{z}}\otimes (G(\theta_{1},0)H-I_{n_{x}})-R'\otimes F(\theta_{1},0)H  \right)^{-1}\times\\
				&&\left((M_{1}Q^{\star}(\theta_{1},\theta_{2}))'\otimes F(\theta_{1},0)\right)vec(H)\end{eqnarray*}
			\vspace{-0.3 in}
		\end{enumerate} \label{repre} \vspace{-0.1 in}
		and $\left(P^{\star}(\theta_{1},\theta_{2}),Q^{\star}(\theta_{1},\theta_{2})\right)$ \normalsize is the Rational Expectations solution to  \eqref{eq:Exp_Sys_Dist}.
\end{prop} \end{spacing} \vspace{-0.2 in} 
\begin{proof}
	See the Appendix
\end{proof} 
This existence result allows us to directly parameterize the distortions ${\mu}_{t}$ in   \eqref{eq:Exp_Sys_Dist} by ${\mu}_{t}= M_{1}X_{t-1} + M_{2}Z_{t}$, solve the system using the conventional $QZ$ decomposition and construct a likelihood function that depends on the structural parameters, as well as $(M_{1},M_{2})$. The cross-equation restrictions that appear in Proposition \ref{repre} imply that there is a tight link between $\theta_{1}$ and $(\theta_{2},vec(M_{1}),vec(M_{2}))$, which is important for identification. 

The assumption that ${\mu}_{t}$ is a linear function of the states is an implicit assumption that the average value of $x_{i,t}$ and $z_{i,t}$ is sufficient to predict distortions. 
 This does not imply that we invoke a representative agent equivalence as $X_{t}$ can possibly include higher moments of the distribution, as shown in the context of the KS model e.g. the cross sectional variance component $\Xi_{t}$. 

 Moreover, since  $\tilde{\mu}_{t}$ is partially determined by the proportion of constrained agents, in equilibrium, a positive probability of someone being constrained in the future will imply an aggregate precautionary effect, captured by $H$. 
 
 A limitation of the above proposition is the assumption of linearity. Nevertheless, most of the recent  literature on computing HA models with aggregate shocks focuses on linear representations which makes it a useful starting point, see e.g. \citet{REITER2009649,NBERw25020,NBERw24138}.  
 
 Based on these results, I next investigate whether the approximation performs well under parameterizations that imply a varying degree of self-insurance, which changes the exposure of households to idiosyncratic risk.
\vspace{-0.1 in}
\section{Evaluating the empirical approximation to the KS economy}
Since employment also depends on the aggregate shock, I redefine  aggregate productivity to include labor as  e.g. $Z_{L,t}:=Z_{t}L^{1-\alpha}_{t}$. The second order approximation to the KS economy yields the following first order conditions \small
\begin{eqnarray}
K_{t+1}&=& K^{\alpha}_{t}\bar{l}^{1-\alpha}Z_{L,t}+(1-\delta)K_{t}-C_{t}\nonumber\\
r^{k}_{t}&=&\alpha K^{\alpha-1}_{t}\bar{l}^{1-\alpha}Z_{L,t}\nonumber\\
Z_{L,t}&=& \left(\left(1-u_{low}\right)^{1-\alpha}Z_{low}\boldsymbol{1}_{low}+\left(1-u_{high}\right)^{1-\alpha}Z_{high}\boldsymbol{1}_{high}\right)\\
C_{t}^{-\omega}& =& \beta\mathbb{E}_{t}(1-\delta+ r^{k}_{t+1})\left(\frac{\Xi_{t+1}}{\Xi_{t}}\right)C^{-\omega}_{t+1}+\mu_{t}\nonumber
\end{eqnarray} 
\normalsize

where $\boldsymbol{1}_{high}$ is equal to one when the productivity level is high.
In order to obtain a model that does not involve a discrete set of jumps for the productivity shock, I also approximate the indicator functions for the state of productivity using a logistic function: $\boldsymbol{1}_{high} \approx \frac{1}{1+e^{-\zeta(Z_{t}-Z_{ss})}}$, where $Z_{ss}$ is the unconditional mean of the two state process and $\zeta$ is large enough such that the continuous process approximates the two state process well.  

Log-linearizing the resulting system around the non-stochastic steady state and utilizing the result in Proposition \ref{repre}, the  approximating model which is estimable using conventional (Bayesian) methods becomes as follows:
\small 

\begin{eqnarray}
\tilde{K}_{t+1}&=& \left(1-\frac{c_{ss}}{k_{ss}}\right)\tilde{K}_{t} + \frac{c_{ss}}{k_{ss}}\tilde{C}_{t}+\zeta k^{\alpha-1}_{ss}\frac{\gamma_{1}}{2}Z_{ss} \tilde{Z}_{t}\nonumber\\
\tilde{r}^{k}_{t}&=& (\alpha-1)\tilde{K}_{t}+\frac{\zeta}{2}\frac{\gamma_{1}}{\gamma_{1}+2\gamma_{0}}Z_{t}\nonumber\\
\tilde{Z}_{t}&=& \rho{Z}_{t-1}+e_{z,t}\nonumber \\
\tilde{C}_{t} & =& \beta(1-\delta+r^{k}_{ss})\mathbb{E}_{t} \left(\tilde{C}_{t+1}-\frac{1}{\omega}(\tilde{\Xi}_{t+1}-\tilde{\Xi}_{t})-\frac{r^{k}_{ss}}{\omega(1-\delta+r^{k}_{ss})}\tilde{r}^{k}_{t}\right)+\tilde{\mu}_{c,t}\nonumber\\
\tilde{\mu}_{c,t}&=&m_{1}\tilde{K}_{t} + m_{2}\tilde{Z}_{t}\nonumber\\
\tilde{\Xi}_{t}&=&\xi_{1}\tilde{K}_{t} + \xi_{2}\tilde{Z}_{t}\nonumber
\end{eqnarray}
\normalsize
where $(m_{1},m_{2},\xi_{1},\xi_{2})$ are the additional parameters\footnote{In Proposition \ref{repre}, $(M_{1},M_{2})$ will depend on $(m_{1},m_{2},\xi_{1},\xi_{2})$. Also, notice that $\tilde{\Xi}$ is proportional to the variance of the consumption ratio, so directly parameterizing the former is with no loss of generality.} and \small \[\frac{c_{ss}}{k_{ss}}=-\delta+k_{ss}^{\alpha-1}(\gamma_{0}+\frac{\gamma_{1}}{2}), \quad r^{k}_{ss}=\alpha k_{ss}^{\alpha-1}(\gamma_{0}+\frac{\gamma_{1}}{2}),\quad \rho = (Z_{high}-Z_{low})(p_{hh}-p_{lh})\frac{\zeta}{2}\] \[\gamma_{0}=\bar{l}^{1-\alpha}\left(1-u_{low}\right)^{1-\alpha}Z_{low},\quad
 \gamma_{1}=\bar{l}^{1-\alpha}\left(\left(1-u_{high}\right)^{1-\alpha}Z_{high}-\left(1-u_{low}\right)^{1-\alpha}Z_{low}\right)\]

\normalsize \vspace{0.2 in}
\subsection*{Estimation Experiments}
 I adopt the same parameterization and solution method as in \citet{MALIAR201042, denhaanjudd} to generate data from this economy\footnote{$\beta=0.99,\gamma=2,\alpha=0.36,\delta=0.025,Z_{high}=1.01,Z_{low}=0.99,\nu = 0.15,\bar{l}=\frac{1}{0.9}, u_{h}=0.04,u_{l} = 0.1$.}.  In particular, I simulate 1400 periods and keep the last 400 as the pseudo-sample. Since there is only one shock, a normally distributed measurement error with variance equal to $1\%$ of the variances is added to the observables. Based on this sample, I  estimate the free parameters of the log-linear approximating model using all variables as observables. Below, I plot the simulated data against their (Kalman) filtered values using the estimated model parameters $(\zeta,m_{1},m_{2},\xi_{1},\xi_{2},\sigma_{Z})$ evaluated at the median  of the parameter draws. In theory, if the approximation error is negligible, the two time series should coincide.

The quality of the approximation is expected to be very high given the "approximate aggregation result" in KS. The Krusell-Smith equilibrium  computes an allocation where the log linear law of motion of the capital stock is sufficient to predict future prices. Nevertheless, as mentioned in their original work, this is not necessarily a self-fulfilling equilibrium, but there is a fundamental reason for which this equilibrium features aggregation. The agents who hold most of the wealth are well insured against idionycratic risk and are expected to behave more like permanent income consumers. The lower the patience and risk aversion, the worse the quality of the approximation is expected to be.

On the other hand, what is less clear a priori is whether the approximation will be able to track the evolution of the cross sectional variance of consumption and whether the dependence of the variance of shocks is going to be well identified. Proposition \ref{repre} illustrates that the cross-equation restrictions can aid identification.  
\paragraph{\textbf{Experiment 1}: Benchmark parameterization $(\omega=2$, $\beta = 0.99$)}
In this experiment the value of  the preference parameters imply highly patient agents with reasonable levels of risk aversion. The approximation performs very well for macroeconomic aggregates, that is consumption and the capital stock. Moreover, the (log) variance of the consumption shares exhibits significant fluctuations, but is mainly driven by the two aggregate states, and thus our representation is sufficient to capture the dynamics. In the Appendix, I present the results from the second experiment under adverse parameterization (less patient, less risk averse households). As expected, since agents are more exposed to idiosyncratic risk, the variance of consumption cannot be perfectly predicted by the aggregate states. Nevertheless macroeconomic aggregates are still predicted to a high degree of accuracy.



\begin{figure}[H]
\includegraphics[scale=0.45]{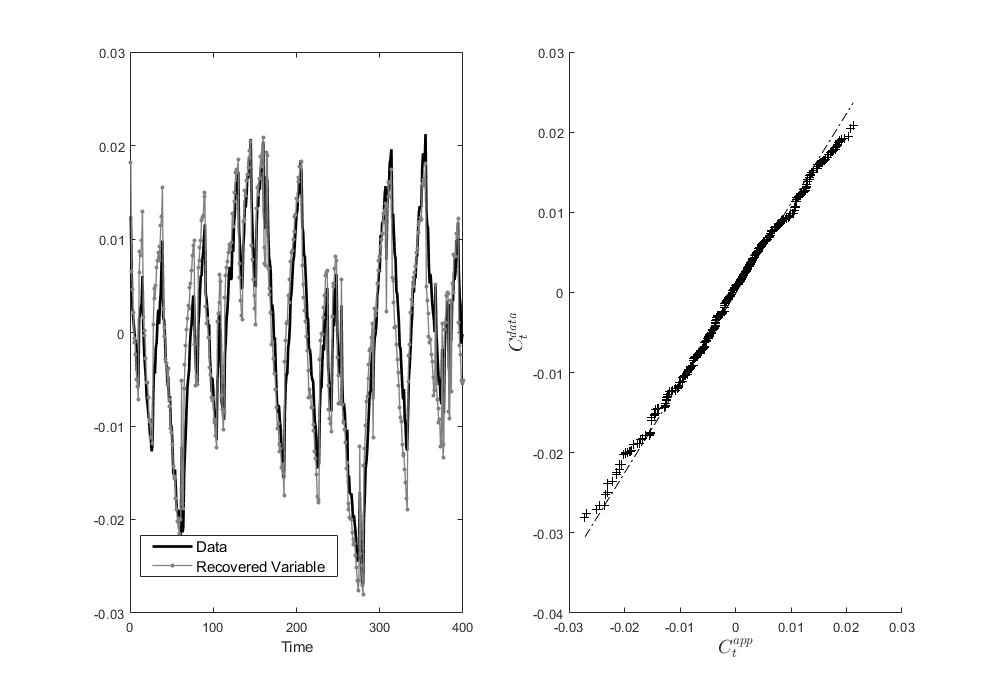}
	\includegraphics[scale=0.45]{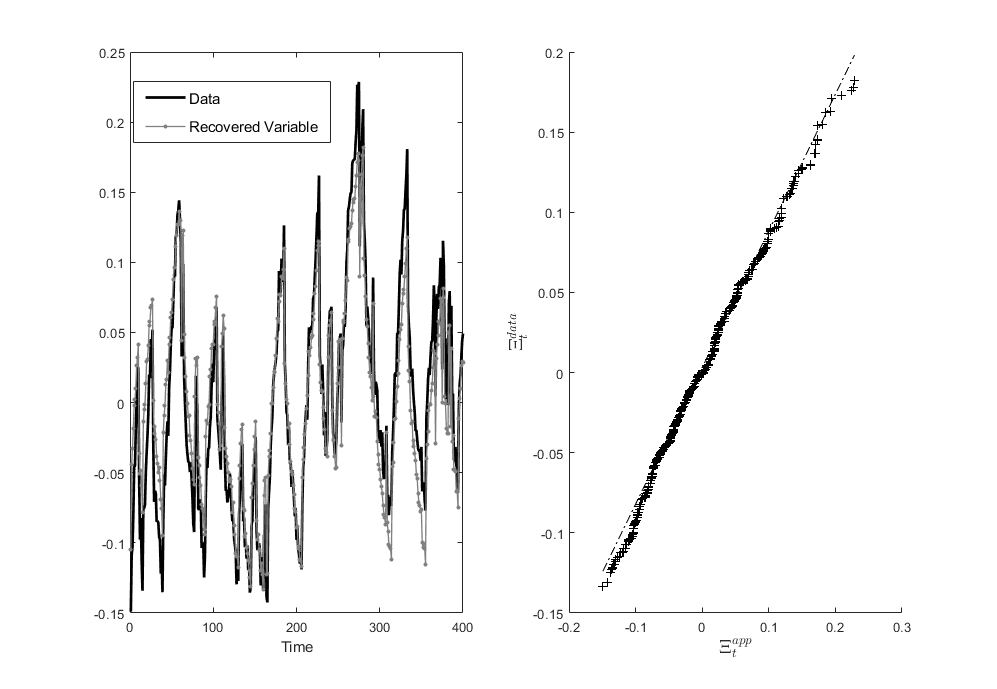}
\vspace{-0.1 in}
\protect\caption{Aggregate Consumption (Upper), Variance of Consumption Shares (Lower)}
\end{figure} 

\begin{figure}[H]

	\includegraphics[scale=0.45]{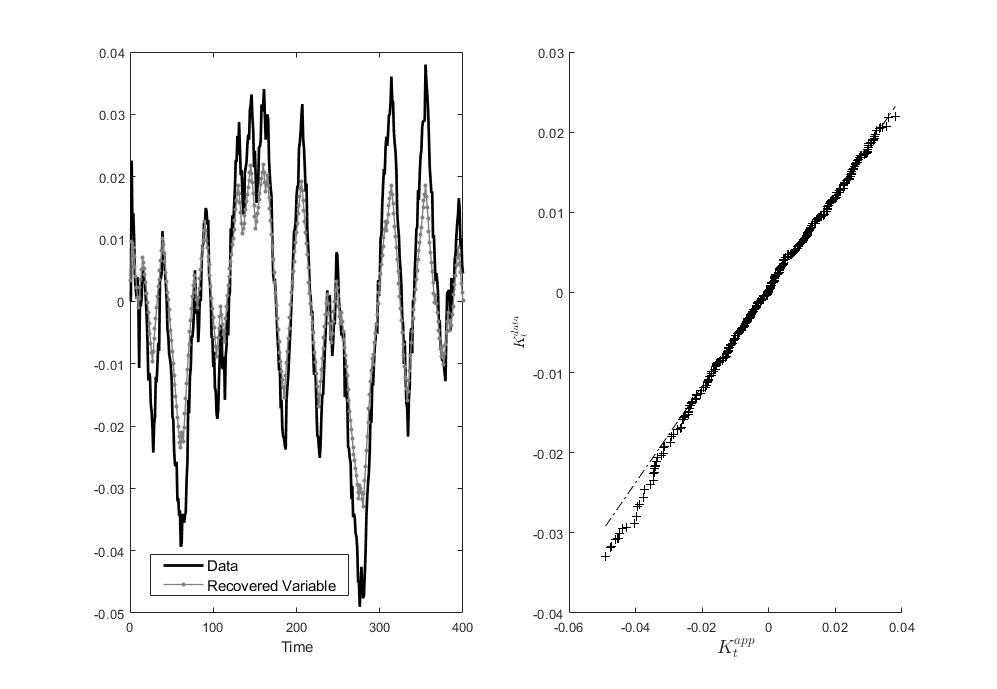}
	\vspace{-0.1 in}
	\protect\caption{Aggregate Capital Stock}
\end{figure} \vspace{-0.2 in}

Having shown that this approximation works well, the next section studies financial frictions in Spain through the lens of a more elaborate HA model informed by aggregate and micro-survey data.

\section{Heterogeneity and Financial Frictions in Spain} 
\begin{figure}[H]
	\begin{centering}
		\includegraphics[scale=0.23]{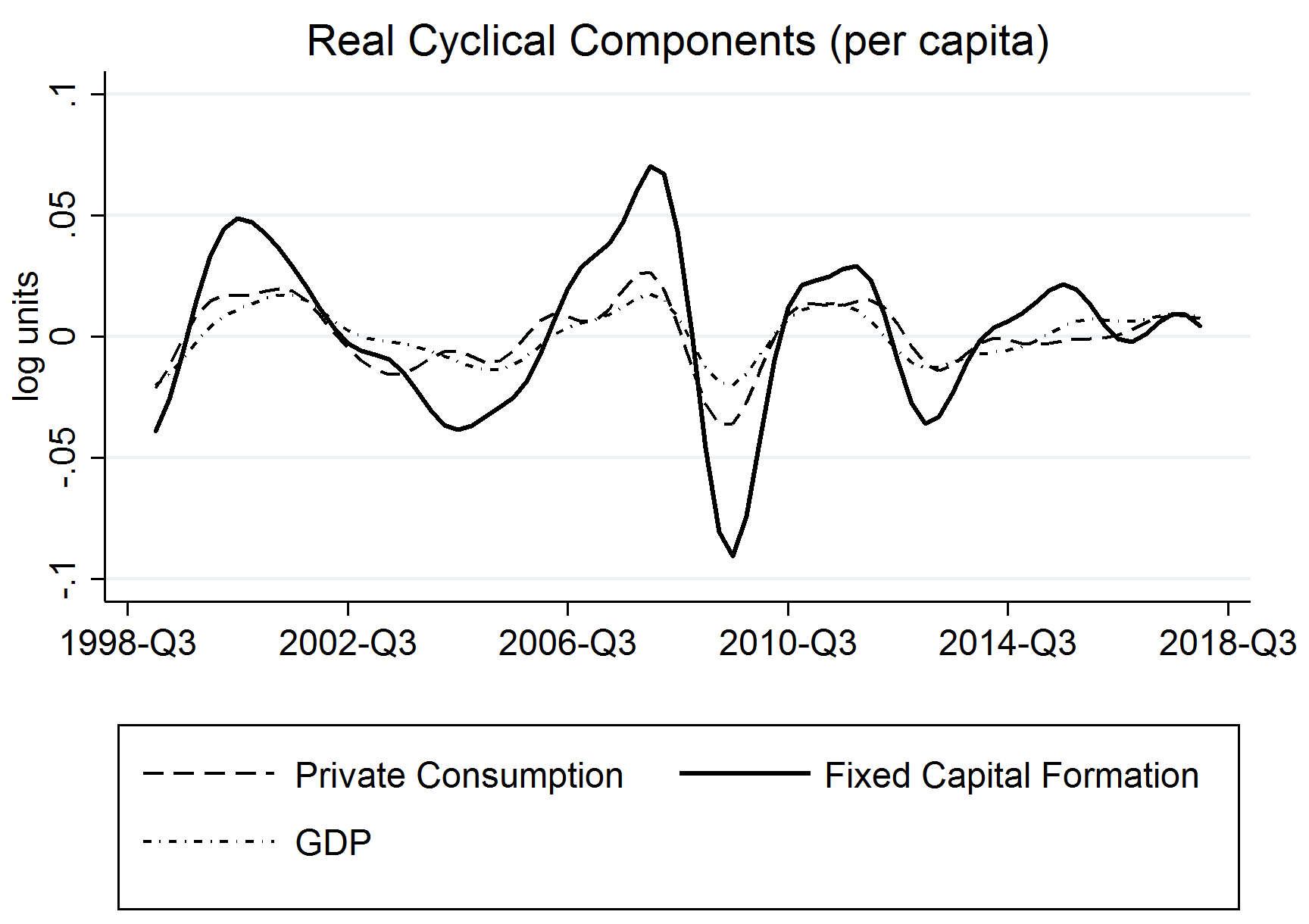}
			\includegraphics[scale=0.42]{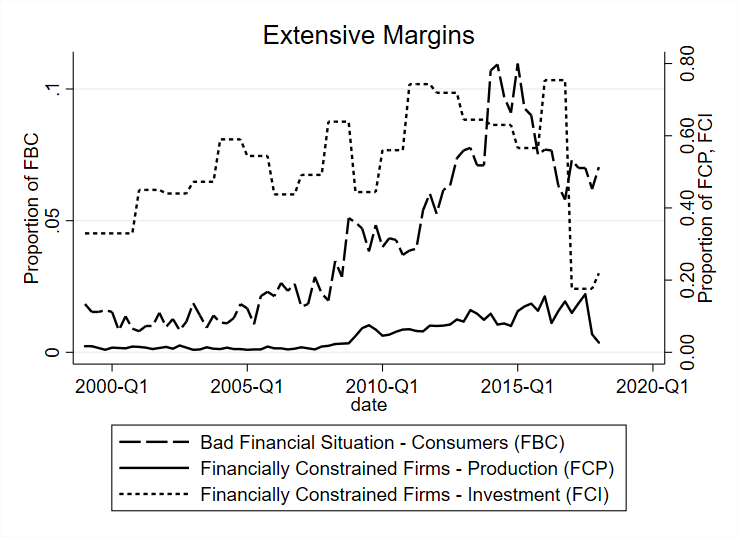}
		\par
	\end{centering}  \vspace{-0.1 in}
	\protect\caption{}
	\label{spain}
\end{figure}

Spain has experienced a severe financial crisis during the period 2008-2014. The initial boom, which was led by unrestrained credit growth and manifested into a housing bubble, was succeeded by a deep recession, a rise in firm bankruptcies, household deleveraging and a large drop in investment. 
Figure \ref{spain} plots the cyclical components for consumption, investment and output after 1999 extracted using the \citet{doi:10.1111/1468-2354.t01-1-00076} filter. 
As evident, there was a large slump in aggregate investment, followed by a relatively smaller slump in aggregate consumption. Similar conclusions are drawn if one looks at the raw series. 

Moreover, a casual observation of aggregated survey data on household's financial position and firm's financial constraints 
reveals the extent to which consumers and firms were affected. 

 \vspace{-0.15 in}
The proportion of industrial firms claiming financial constraints in production (FCP, quarterly) increased from $2\%$ in 2008 to almost $15\%$ before the end of 2015\footnote{BCS data can be downloaded from \href{https://ec.europa.eu/info/business-economy-euro/indicators-statistics/economic-databases/business-and-consumer-surveys_en}{here}.}. The proportion of financially strained households (those that claim to be getting indebted, (FBC, monthly)) increased from $5\%$ in 2008 to $10\%$ by the end of 2016. Although more volatile, the share of firms claiming lack of financial resources for investment (FCI, bi-annual)  developed an upward trend from 2008, with a sharp reversal after the end of 2016 where it reached $75\%$\footnote{Please see the Appendix for the exact wording of the question and data transformations.}. 

These observations beg the question of what is the corresponding role of financial frictions in explaining business cycle fluctuations in Spain, and what is their interaction with aggregate shocks. I next exploit the theoretical analysis of the previous sections by building a more elaborate quasi-structural model to provide some empirical evidence to bear on this question.

I will proceed by gradually building the empirical model that is going to be used in estimation. First, I will introduce the benchmark heterogeneous agent model in its frictionless case where I will discuss the complications that arise when we consider aggregation and how our approach is robust to alternative mechanisms that could generate aggregate distortions.  I will then add some additional features such as variable capacity utilization, which generates additional cross sectional moments, price markups and government policy. \vspace{-0.1 in}
\subsection{The Benchmark Heterogeneous Agent Model} 
Since financial frictions can manifest themselves through the household or the firm side, I consider a two asset economy where households can save in capital as well as in bonds. This is the minimal structure that is relevant to illustrate the empirical restrictions implied by financial constraints in general equilibrium and the cross sectional moments that naturally arise when we approximate the HA model around the representative agent allocation. A more elaborate model will in principle generate more cross sectional moments that will have to be identified from micro-surveys and linked to the business cycle model. The frictionless model is presented first, and it is demonstrated that heterogeneity is not of first order importance. However, this will not be true in an economy with frictions. For brevity of notation, let $\Lambda_{t}(i)$ signify the cross sectional distribution of agents across state variables at time $t$\footnote{$\Lambda_{t}(i)$ denotes the joint distribution function at time $t$, which is $\mathbb{P}_{t}(z_{i,t}\leq z,k_{i,t}\leq k)$ in this section. This is equivalent to $p(s_{i,t}|S_{t})$ in the previous sections.}, and denote aggregate variables by capital letters i.e. $X_{t}\equiv \int {x_{i,t}d\Lambda_{t}(i)}$. Similarly, denote  by $\Lambda_{t,t+1}(i)$ the joint distribution of individual states at $(t,t+1)$. 
\subsubsection{The Frictionless Model} 
On the supply side, households with positive capital holdings turn into effective entrepreneurs, who are akin to intermediate producers facing a downward sloping demand curve. Moreover, aggregate output is a CES aggregate of intermediate goods. 

Under monopolistic competition in the intermediate goods sector, the supply side has a standard formulation.    The final good firm chooses aggregate and firm specific output $\left(Y_{{t},}y_{i,t}\right)$ to maximize profits $P_{t}Y_{t}-\int p_{i,t}y_{i,t}d\Lambda_{t}(i)$ subject to CES aggregation, $Y_{t}^{\frac{\epsilon-1}{\epsilon}}=\int y_{i,t}^{\frac{\epsilon-1}{\epsilon}}d\Lambda_{t}(i)$, leading to a downward sloping demand curve for each firm $i$, $y_{i,t}=\left(\frac{p_{i,t}}{P_{t}}\right)^{-\epsilon}Y_{t}$. 

The effective entrepreneur ($k_{i,t}>0$) hires labor $(l^{d}_{i,t})$ to minimize the cost of production $W_{t}l^{d}_{i,t}$ subject to producing a certain level of output $(z_{i,t}k_{i,t})^{\alpha}(l^{d}_{i,t})^{1-\alpha}\geq y_{i,t}$. This yields nominal labor costs equal to  $W_{t}l^{d}_{i,t}=(1-\alpha)\theta_{i}y_{i,t}$ where $\theta_{i}$ is the multiplier on the output constraint and $(1-\alpha)\theta_{i}$ is the marginal cost. 

Correspondingly, output at the optimal choice of labor is equal to $y_{i,t}=z_{i}k_{i}\theta_{i,t}^{\frac{1-\alpha}{\alpha}}\zeta_{t}$  where $\zeta_{t}\equiv\left(\frac{1-\alpha}{W_{t}}\right)^{\frac{1-\alpha}{\alpha}}$. Under the optimal price (constant markup over marginal cost), $p_{i,t}=\frac{\epsilon}{\epsilon-1}(1-\alpha)\theta_{i,t}$, and therefore indirect profit is equal to $pr_{i,t}=\frac{y_{i,t}p_{i,t}}{\epsilon}$. Thus, aggregate labor costs are a fraction of nominal GDP:  \begin{eqnarray*}W_{t}L_{t}=\frac{\epsilon-1}{\epsilon}P_{t}Y_{t}\end{eqnarray*}\vspace{-0.25 in}
\paragraph*{\textit{Households.}}
Dynastic households decide over consumption, working hours and total net worth to be carried over to the next period, $(c_{i,t},l^{s}_{i,t}, a_{i,t+1})$. In turn, $a_{i,t}$ is split between bonds $w_{i,t}$ earning a gross nominal interest rate $R_{t}$ and capital $k_{i,t}$, earning a gross return $R^{k}_{i,t}$. 

Individual investment decisions $(\iota_{i,t})$ increase the availability of capital for next period, up to a certain level of depreciation. 
Gross individual income is thus equal to $\bar{y}_{i,t}= W_{t}l_{i,t} + (1-\delta)k_{i,t} + pr_{i,t}$. 

If $k_{i,t}=0$, the household-worker receives only labor income. The household's decision problem is therefore as follows: \begin{spacing}{1.5}
	\begin{eqnarray*}
		&&\max_{\{y_{i,t},k_{i,t+1},w_{i,t+1},\iota_{i,t},c_{i,t}\}_{1}^{\infty }}\mathbb{E}_{0}\sum_{t=0}^{\infty }\beta
		^{t}\left(\frac{c_{i,t}^{1-\omega }-1}{1-\omega }- \frac{l_{i,t}^{1+\eta}}{1+\eta}  \right)\\
		s.t.  &a_{i,t+1} &= R_{t}w_{i,t} + \bar{y}_{i,t}-P_{t}c_{i,t}\\
		&a_{i,t+1} & = P_{t}k_{i,t+1} + w_{i,t+1}\\
		&\bar{y}_{i,t} &= W_{t}l_{i,t} + (1-\delta)P_{t}k_{i,t} + pr_{i,t}\\
		&k_{i,t+1} &=(1-\delta )k_{i,t}+  \iota_{i,t}\geq 0
		\end{eqnarray*} \vspace{-0.15 in}
	Optimal behavior is then characterized by the following first order conditions: \vspace{-0.25 in}
	\begin{eqnarray*}
		c^{-\omega}_{i,t}&=&\beta\mathbb{E}_{t}c^{-\omega}_{i,t+1}R^{k}_{i,t+1} + v_{i,t}\\
		c^{-\omega}_{i,t}&=&\beta\mathbb{E}_{t}c^{-\omega}_{i,t+1}\Pi^{-1}_{t+1}R_{t+1}\\
		l^{\eta}_{i,t} &=& \frac{W_{t}}{P_{t}}c^{-\omega}_{i,t}\\
		v_{i,t}k_{i,t+1}&=& 0 
\end{eqnarray*} \end{spacing} \vspace{-0.1 in}
which implies that $
v_{i,t}=\beta\mathbb{E}_{t}c^{-\omega}_{i,t+1}\left(\Pi^{-1}_{t+1}R_{t+1}-R^{k}_{i,t+1}\right)$
where $v_{i,t}$ is the multiplier on the non-negativity constraint on $k_{i,t}$ and $\Pi_{t+1}=\frac{P_{t+1}}{P_{t}}$ is gross inflation and $R^{k}_{i,t+1}=(1-\delta) + \frac{\partial pr_{i,t+1}}{\partial k_{i,t+1}}\frac{1}{P_{t+1}}$, the individual return to capital investment.

If $z_{i,t}$ is highly autocorrelated, then bad productivity realizations can induce the household to hit the non-negativity constraint where it is optimal to not hold any capital stock $(v_{i,t}>0)$. Note also that there is no assumed borrowing limit (apart from the natural limit below which nobody would like to be) and thus workers can effectively save to self insure against idiosyncratic risk. 
\subsubsection{The complete markets case}
As in the KS example, growth in marginal utility is equalized across agents. This implies that individual consumption is proportional to aggregate consumption and the representative agent approximation becomes exact: \vspace{-0.1 in}
\begin{eqnarray*}
	C^{-\omega}_{t}&=&\beta\mathbb{E}_{t}C^{-\omega}_{t+1}\Pi^{-1}_{t+1}R_{t+1}
\end{eqnarray*} 

which also implies that cross sectional variation becomes time invariant:\small
\[\int c^{-\omega}_{i,t} \Lambda_{t}(i) =  C^{-\omega}_{t}\left(1+ \frac{\omega(\omega+1)}{2}\int(c_{i}-1)^{2}d\Lambda_{t}(i) + o(1)\right)   \] \normalsize

Regarding the first order condition for capital, multiplying with predetermined $\frac{k_{i,t+1}}{K_{t+1}}$, applying the same approximation to marginal utility and aggregating using $\Lambda_{t+1}(i)$ yields that\footnote{The same result holds if we apply the 2nd order approximation, integrate using the joint distribution $\mathbb{P}_{t}(z_{i,t}\leq z,k_{i,t}\leq k)$ and set higher order moments to zero.}:  \small
\begin{eqnarray*}
	C^{-\omega}_{t}&=&\beta\mathbb{E}_{t}C^{-\omega}_{t+1}\int R^{k}_{i,t+1}\frac{k_{i,t+1}}{K_{t+1}} d\Lambda_{t+1}(i)= \beta\mathbb{E}_{t}C^{-\omega}_{t+1}\left(	(1-\delta)+ \frac{\alpha}{\epsilon}\frac{Y_{t+1}}{K_{t+1}}\right)\end{eqnarray*} \normalsize
When $\epsilon\to 1$, this condition is isomorphic to the representative agent case where aggregate production is Cobb- Douglas and $\alpha \frac{Y_{t+1}}{K_{t+1}}$ is the marginal product of capital. 
Furthermore, integrating firm revenue, and solving for aggregate output yields: \small
\[Y_{t}= \left(\int z_{i}k_{i}\left(\frac{p_{i,t}}{P_{t}}\right)^{\frac{1}{\alpha}}d\Lambda_{t}(i)\right)^{\alpha}L_{t}^{1-\alpha}\] \normalsize  
A second order approximation yields  $Y_{t}\approx \left(Z_{t}K_{t}(1+pd_{t})\right)^{\alpha}L^{1-\alpha}_{t}$, where \small \[pd_{t}:=\frac{1}{\alpha}\int \frac{z_{i,t}}{Z_{t}}\frac{k_{i,t}}{K_{t}}\left(\frac{p_{i,t}}{P_{t}}-1\right)^{2}d\Lambda_{t}(i)\] \normalsize is a size and productivity weighted measure of price dispersion. 
Since in most applications, $Z_{t}$ and $K_{t}$ are unobserved, unless price dispersion is time invariant, recovered estimates of productivity shocks and the capital stock will also reflect changes that are attributed to products that become relatively less or more expensive, affecting output on the margin. 

Finally, the inter-temporal condition for hours worked under the same approximation for the marginal utility of consumption yields that aggregate labor supply is pinned down by
 $L_{t}=\left(\frac{W_{t}}{P{t}}\right)^{\frac{1}{\eta}}C^{-\frac{\omega}{\eta}}_{t}$ while in equilibrium  \small
  \[\frac{\epsilon-1}{\epsilon}Y_{t}=\left(\frac{W_{t}}{P{t}}\right)^{1+\frac{1}{\eta}}C^{-\frac{\omega}{\eta}}_{t}\] \normalsize
Denoting by $\tilde{X}$ the percentage deviations of $X_{t}$ from the steady state, the log-linearized conditions are identical to the representative agent case. Heterogeneity is thus negligible to first order: \vspace{-0.4 in}\begin{spacing}{1.4}
	\begin{eqnarray*}
		-\omega \tilde{C}_{t}-{\mathbb{E}_{t}}\tilde{R}_{t+1}+\omega {\mathbb{E}_{t}}\tilde{C}_{t+1} &=&0 \\
		-\omega \tilde{C}_{t}-(1-\beta(1-\delta)){\mathbb{E}_{t}}\tilde{R}^{k}%
		_{t+1}+\omega {\mathbb{E}_{t}}\tilde{C}_{t+1} &=&0 \\
		y_{ss}\tilde{Y_{t}}-c_{ss}\tilde{C}_{t}-i_{ss}\tilde{I}_{t} &=&0 \\
		\tilde{K}_{t+1}-(1-\delta )\tilde{K}_{t}-\delta \tilde{I}_{t} &=&0 \\
		\tilde{Y}_{t}-\alpha\tilde{Z}_{t}-\alpha \tilde{K}_{t}- (1-\alpha)\tilde{L}_{t} &=&0 \\
		\tilde{R^{k}_{t}}-\alpha\tilde{Z}_{t}+(1-\alpha )\tilde{K}_{t} -(1-\alpha)\tilde{L}_{t} &=&0 \\
		\tilde{Y}_{t} - \tilde{L}_{t}-\omega \tilde{C}_{t}-\eta\tilde{L}_{t} &=& 0
\end{eqnarray*} \end{spacing}\vspace{-0.2 in}
\subsubsection{When is heterogeneity important for aggregate fluctuations?}$ $
This section illustrates that in the presence of frictions, heterogeneity matters to first order.  More particularly, in addition to productivity,  this section considers another dimension of heterogeneity, that of investment efficiency. Both productivity and investment efficiency shocks, $z_{i,t}$ and $\eta_{i,t}$ respectively, can potentially affect aggregate consumption and investment through different channels. \citet{buera_moll} studied these channels in isolation and concluded that heterogeneity in productivity is the most likely form of heterogeneity as it leads to a fall in total factor productivity (TFP), as is often observed in recessions. 

Nevertheless,  \citet{10.1093/qje/qjx024} (GKKV henceforth) have argued that models of financial frictions that link TFP to financial frictions predict an increase in TFP after a financial liberalization episode, which is inconsistent with the experience of Southern Europe, including Spain. 

The benchmark model is next modified to characterize the interaction of these shocks and the borrowing constraint on net worth\footnote{The way the borrowing constraint is specified does not affect the qualitative statements made for the distortions ; $\varphi_{i,t}$ could in principle depend on the size of the firm $k_{i,t+1}$ as in GKKV or the market value of its capital stock, $\eta_{i,t}$.}. 

The household problem becomes as follows: 
\begin{eqnarray*}
	&&\max_{\{y_{i,t},k_{i,t+1},w_{i,t+1},\iota_{i,t}\}_{1}^{\infty }}\mathbb{E}_{0}\sum_{t=0}^{\infty }\beta
	^{t}\left(\frac{c_{i,t}^{1-\omega }-1}{1-\omega }- \frac{l_{i,t}^{1+\eta}}{1+\eta}  \right)\\
	s.t.  &a_{i,t+1} &= R_{t}w_{i,t} + \bar{y}_{i,t}-P_{t}c_{i,t}\\
	&a_{i,t+1} & = P_{t}\eta_{i,t}k_{i,t+1} + w_{i,t+1}\\
	&\bar{y}_{i,t} &= W_{t}l_{i,t} + (1-\delta)P_{t}\eta_{i,t}k_{i,t} + pr_{i,t}\\
	&k_{i,t+1} &=(1-\delta )k_{i,t}+  \frac{\iota_{i,t}}{\eta_{i,t}}\\
	&k_{i,t+1} & \geq 0 \\
	& {w_{i,t+1}}& \geq-{P_{t}}\varphi_{i,t} k_{i,t+1}
\end{eqnarray*}   
and the corresponding first order conditions with respect to capital and bonds are: \vspace{-0.2 in}
\begin{eqnarray*}
	\eta_{i,t}c^{-\omega}_{i,t}&=&\beta\mathbb{E}_{t}c^{-\omega}_{i,t+1}R^{k}_{i,t+1} + v_{i,t} +\varphi_{i,t}\mu_{i,t}\\
	c^{-\omega}_{i,t}&=&\beta\mathbb{E}_{t}c^{-\omega}_{i,t+1}\Pi^{-1}_{t+1}R_{t+1} + \mu_{i,t}
\end{eqnarray*} 
where $\mu_{i,t}$ is the multiplier on the occasionally binding borrowing constraint.
Combining the two conditions, $\mu_{i,t}$ when non zero is equal to \[\mu_{i,t}=\beta\mathbb{E}_{t}c^{-\omega}_{i,t+1}\left(\frac{R^{k}_{i,t+1}-\eta_{i,t}\Pi^{-1}_{t+1}R_{t+1}}{\eta_{i,t}-\varphi_{i,t}}\right)+\frac{v_{i,t}}{\eta_{i,t}-\varphi_{i,t}}\]  Since borrowing constraints can bind, households are not perfectly insured, resulting in a non-negligible consumption variance.
\subsubsection{Aggregation}
Following the same approach as before, 
the approximate aggregated first order condition for bonds is equal to:
\begin{eqnarray}
\Xi_{t} C^{-\omega}_{t}&=&\beta\mathbb{E}_{t}C^{-\omega}_{t+1}{\Xi_{t+1}}{}\Pi^{-1}_{t+1}R_{t+1} + \int \mu_{i,t} d\Lambda_{t}(i)  \label{distR}
\end{eqnarray}
Correspondingly, the approximate aggregate investment Euler equation is
\begin{eqnarray}
\Xi_{t} C^{-\omega}_{t}&=&\beta\mathbb{E}_{t}C^{-\omega}_{t+1}{\Xi_{t+1}}{}\int R^{k}_{i,t+1}d\Lambda_{t}(i) + \int \mu^{k}_{i,t} d\Lambda_{t}(i) \label{distRk}
\end{eqnarray}
where $\mu^{k}_{i,t}=\beta\mathbb{E}_{t}c^{-\omega}_{i,t+1}\left(\frac{(1-\eta_{i,t}+\varphi_{i,t})R^{k}_{i,t+1}-\Pi^{-1}_{t+1}R_{t+1}}{\eta_{i,t}-\varphi_{i,t}}\right)+\frac{v_{i,t}}{\eta_{i,t}-\varphi_{i,t}}$.

If $\eta_{i,t}=\varphi_{i,t}=1$, the aggregate distortions to the bond and capital Euler equations are identical, which implies that the aggregate expected returns are equalized as in the standard complete markets case. If not, then the distortions will be different, causing a disconnect in the aggregate returns, while distortions will generally  depend on aggregate shocks in a different way. An alternative mechanism would imply a different dependence.This provides further justification to our semi-structural approach that aims to measure the distortions and how they depend on aggregate shocks as compared to imposing a particular structure. This dependence will be empirically identified.

%


\section{Full Model: A Richer Specification}
Having analyzed the main implications of heterogeneity for macroeconomic aggregates, and with a view to taking the model to the data, I enrich the model on several dimensions. 

I particular, household utility is assumed to also depend on external habits while households decide also on capacity utilization, with the corresponding cost of utilization being quadratic: $\frac{\psi}{2}u^{2}_{i,t}$. Capital holdings are thus also affected by an intensive margin e.g. by how much capital is utilized. 

Therefore, indirect profit returns become \[\frac{\partial pr_{i,t}}{\partial k_{i,t}} =  \frac{\partial y_{i,t}}{\partial k^{u}_{i,t}}\frac{p_{i,t}}{\epsilon_{t}}{u}_{i,t}-\frac{\psi}{2}u^{2}_{i,t}\] with utilized capital, $k^{u}_{i,t}= k_{i,t}u_{i,t}$, and a time varying price markup, $\nu^{p}_{t}:=\frac{\epsilon_{t}}{\epsilon_{t}-1}$. 

As for government policy, I introduce  tax rates $\tau^{inc}_{t}$ on wage income and  $\tau^{corp}_{t}$ on net profits respectively and lump sum taxes (net of transfers), $t_{i}$. 
The household problem is therefore almost identical to the previous section, with  cash on hand being the only difference:
\begin{eqnarray*}
	&&\max_{\{y_{i,t},k_{i,t+1},w_{i,t+1},\iota_{i,t},u_{i,t}\}_{1}^{\infty }}\mathbb{E}_{0}\sum_{t=0}^{\infty }\beta
	^{t}\left(\frac{\left(c_{i,t}-hC_{t-1}\right)^{1-\omega }-1}{1-\omega }- \frac{l_{i,t}^{1+\eta}}{1+\eta}  \right)\\
	s.t.  &a_{i,t+1} &= R_{t}w_{i,t} + \bar{y}_{i,t}-P_{t}c_{i,t}\\
	&a_{i,t+1} & = P_{t}\eta_{i,t}k_{i,t+1} + w_{i,t+1}\\
&\bar{y}_{i,t} =& W_{t}l_{i,t}(1-\tau^{inc}_{t}) + (1-\delta)P_{t}\eta_{i,t}k_{i,t} + (1-\tau^{corp}_{t})\left(pr_{i,t}-P_{t}\frac{\psi}{2}u^{2}_{i,t}k_{i,t}\right)-t_{i}\\
	&k_{i,t+1} &=(1-\delta )\eta_{i,t}k_{i,t}+  \frac{\iota_{i,t}}{\eta_{i,t}}\\
	&k_{i,t+1} & \geq 0 \\
	& {w_{i,t+1}}& \geq-{P_{t}}\varphi_{i,t} k_{i,t+1}\\
	\end{eqnarray*}
The optimality conditions also remain the same, with the only difference that capital returns are now equal to
\begin{eqnarray*}
R^{k}_{i,t} &=& (1-\delta)\eta_{i,t} + (1-\tau^{corp}_{t}) \left(u_{i,t}\frac{\partial y_{i,t}}{\partial k^{u}_{i,t}}\frac{p_{i,t}}{P_{t}} \frac{\nu^{p}_{t}-1}{\nu^{p}_{t}} -\frac{\psi}{2}u^{2}_{i,t}    \right)\end{eqnarray*} 
and optimal capital utilization pinned down by: 
\begin{eqnarray} u_{i,t} &=& \frac{1}{\psi}\left(\frac{\partial y_{i,t}}{\partial k^{u}_{i,t}}\frac{p_{i,t}}{P_{t}}\frac{\nu^{p}_{t}-1}{\nu^{p}_{t}}\right) \label{capu}\end{eqnarray} which implies that returns simplify to
$R^{k}_{i,t}= (1-\delta)\eta_{i,t} + (1-\tau_{t}) \frac{\psi}{2}u^{2}_{i,t}$.

In addition, the intratemporal condition is distorted by the tax rate, that is, disposable wage is now $w_{t}(1-\tau^{inc}_{t})$. 
\subsection{Government Budget and Fiscal Policy}
The government finances its primary deficit either through domestic or foreign bond markers. Denoting by $D_{t}$ as the foreign portfolio, the government budget constraint is: 
\[D_{t+1}+\frac{\mathbb{W}_{t+1}}{P_{t}}+Tr_{t} = G_{t}+R_{t}\frac{\mathbb{W}_{t}}{P_{t}}+R^{\star}_{t}D_{t}\] 

where $R^{\star}_{t}$ is the interest rate on international borrowing which can in principle be different than the monetary policy rate due to a country specific premium.
Moreover, it is assumed that the government implements a \textit{tax revenue target} rule, which responds to domestic and international real debt with a potentially different weight:\vspace{-0.1 in} 
\[Tr_{t} = \phi_{f}\left(D_{t+1}-R^{\star}_{t}D_{t}\right)+\phi_{d}\left(\frac{\mathbb{W}_{t+1}}{P_{t}}-R_{t}\frac{\mathbb{W}_{t}}{P_{t}}\right)+\epsilon_{Tr,t}\]
while tax revenue can be generated using marginal income and corporate tax rates and lump sum taxes (net of transfers): \vspace{-0.05 in}
 \begin{eqnarray*}Tr_{t}&=&\tau^{inc}_{t}W_{t}\int l_{i,t}d\Lambda_{t}(i)+\tau^{corp}_{t}\int pr_{i,t}d\Lambda_{t}(i)+T^{lump}\\&=&\tau^{inc}_{t}W_{t}L_{t}+\tau^{corp}_{t}{Y_{t}}\frac{\nu^{p}_{t}-1}{\nu^{p}_{t}}+T^{lump}\end{eqnarray*}
This endogenizes tax decisions, while it incorporates a more realistic component of tax policy in the recent years, that of generating tax revenue to meet a fiscal target. 

Finally since the tax rate portfolio is undetermined, I assume that income tax and corporate tax rates differ by a deterministic and an exogenous iid component, broadly in line with empirical evidence (See the Appendix, Figure \ref{taxrates}.). Thus, I set $\tau^{inc}_{t}= \tau_{1}\tau^{corp}_{t}\epsilon_{d\tau,t}$.

\subsection{Ex Ante Real Rate}

In this small open economy setup, the nominal rate  is largely determined exogenously $(R_{t+1}=R^{\star\star}_{t})$\footnote{Consistent with previous setup, the nominal interest applies to beginning of period bonds, and is thus known from time $t$.}. The ex- ante real rate will therefore be equal to:
\begin{eqnarray}
    r_{t+1} &=& \mathbb{E}_{t}\frac{1+r^{\star\star}_{t}}{1+\pi_{t+1}}-1 \label{eq:exanter}
\end{eqnarray}
While   $r^{\star\star}_{t}$ is mainly influenced by short term and long term foreign factors, expectations about future domestic inflation will in general depend on domestic as well as foreign factors. Instead of hard-wiring the dependence of inflation on these factors through a particular mechanism e.g. a New Keynesian Phillips curve, I will allow for a flexible specification that will depend on a transitory component that is unrelated to domestic fundamentals and the trend component that summarizes the permanent effects of (domestic) aggregate shocks e.g.  productivity. The latter will explain the very persistent portion of variation in inflation (and hence inflation expectations) that can also generate persistent differences between the home country (i.e. Spain) and the rest of the monetary union. To see this suppose that the EMU wide Taylor rule depends on some form of efficient rate, EMU wide inflation expectations and a monetary policy shock. Then, the log-linearized form of  \ref{eq:exanter} is as follows
\begin{eqnarray*}
\tilde{r}_{t+1} &=& \tilde{\rho}^{\star\star}_{t} + \mathbb{E}_{t}\tilde\pi^{\star\star}_{t+1}-\mathbb{E}_{t}\tilde\pi_{t+1} + \epsilon_{r^{\star\star},t}
\end{eqnarray*}
where the first three terms are more likely to depend on both short and long-term structural factors.

\subsection{Aggregation and Equilibrium}
Before aggregating up the economy, it is important to understand that the quasi-structural aggregate model that will be taken to the data will not adhere to the specific mechanism that generates the distortions to consumption and investment, in the same way in which we did not adhere to the specific mechanism in the KS model.

\subsubsection{\textit{Aggregate Investment}.}
Adjusting the earlier aggregate Euler equations for bonds and capital with external habits:
\begin{eqnarray}
	\Xi_{t} (C_{t}-hC_{t-1})^{-\omega}&=&\beta\mathbb{E}_{t}(C_{t+1}-hC_{t})^{-\omega}\Pi^{-1}_{t+1}{\Xi_{t,t+1}}{}R_{t+1} + \mu^{c}_{t} \nonumber\\
	\Xi_{t} (C_{t}-hC_{t-1})^{-\omega}&=&\beta\mathbb{E}_{t}(C_{t+1}-hC_{t})^{-\omega}{\Xi_{t,t+1}}{} R^{k}_{t+1} + \mu^{k}_{t}  \label{eq:invEuler}
\end{eqnarray}
Since the real interest rate is exogenous due to the small open economy assumption, aggregate consumption is already pinned down by the consumption Euler equation, which implies that in the absence of ad-hoc investment adjustment constraints $(\text{e.g:}\quad\eta_{i,t}=\varphi_{i,t}=1)$, investment is residually determined. 

 As $(\mu^{c}_{t},\mu^{k}_{t})$ are not be pinned down by a specific mechanism, investment will not adhere to a specific mechanism either. Instead of exploiting \eqref{eq:invEuler}, distortions to investment arising from e.g. efficiency losses in the investment sector can be modeled directly through the capital accumulation equation\footnote{In the Appendix, I discuss how the model change if we also incorporated the capital Euler equation and how distortions in investment 
 	$\lambda_{I,t}$ depend on the aggregate value of capital. This of course increases the number of parameters that have to be identified.}.
 To see this, notice that if investment is distorted by an idiosyncratic shock,  $k_{i,t+1}=(1-\delta)k_{i,t}+\frac{\iota_{i,t}}{\eta_{i,t}}$. Aggregating, we get that  \small
\[K_{t+1}=(1-\delta)K_{t}+ I_{t} - \int \iota_{i,t}\frac{\eta_{i,t}-1}{\eta_{i,t}}d\Lambda_{t}(i)=(1-\delta)K_{t}+ I_{t} - \lambda_{I,t}\] \normalsize where $\iota_{i,t}\equiv\iota(s_{i,t},S_{t})$ in equilibrium. As long as some agents face investment costs e.g. $\eta_{i,t}>1$, investment is distorted downwards, and the distortion will depend on aggregate shocks through $\iota_{i,t}$\footnote{As noted by \citet{buera_moll}, several papers modeled financial frictions as a result of heterogeneity in investment costs (opportunities), see e.g. \cite{Wang2012,10.1086/262072} .}. In equilibrium, investment will be distorted by both the savings decisions of the households (summarized by $Var_{t}(c_{i,t})$ and $\mu_{t}$) and the distortion $\lambda_{I,t}$, which again, does not adhere to a specific mechanism. 

\subsubsection{\textit{Capital Utilization}.} Empirical measures of e.g. the capital utilization rate and the marginal product of capital across firms are usually size-weighted. To be consistent with these measurements, I re-express \eqref{capu} in terms of the firm capital stock:  \begin{eqnarray} u^{2}_{i,t} &=& \frac{\alpha}{\psi}\frac{y_{i,t}}{ k_{i,t}}\frac{p_{i,t}}{P_{t}}\frac{\nu^{p}_{t}-1}{\nu^{p}_{t}} \label{capu2}\end{eqnarray} 
Moreover, to second order, the size weighted aggregate capital utilization is:
\[U_{t}\equiv \int u_{i,t}\frac{k_{i,t}}{K_{t}}d\Lambda_{t}(i)= \left(\frac{1}{\psi}\frac{\nu^{p}_{t}-1}{\nu^{p}_{t}}\right)^{\frac{1}{2}}\left(\alpha\frac{Y_{t}}{K_{t}}\right)^{\frac{1}{2}}\left(1-\frac{1}{8}V^{k}_{t}\right)\] 
 where $V^{k}_{t}$ is the cross sectional dispersion of the real marginal product across firms\footnote{The aggregate return to capital is also pinned down as follows: 
 	\[R^{k}_{t}= \int R^{k}_{i,t}\frac{k_{i,t}}{K_{t}} d\Lambda_{i}(t)= (1-\delta)\int \eta_{i,t}\frac{k_{i,t}}{K_{t}} d\Lambda_{i}(t) + (1-\tau_{t}) \left(\frac{\psi}{2}\alpha \frac{Y_{t}}{K_{t}}\frac{\nu^{p}_{t}-1}{\nu^{p}_{t}}\right)\]}: \small
\[V^{k}_{t}:=\left(\alpha\frac{Y_{t}}{K_{t}}\right)^{-2}\int \left(\frac{\partial y_{i,t}}{\partial k_{i,t}}\frac{p_{i,t}}{P_{t}}-\alpha\frac{Y_{t}}{K_{t}}\right)^2\frac{k_{i,1}}{K_{t}}d\Lambda_{t}(i)\] \normalsize
Dispersion in the marginal product of firms will lead to lower aggregate capital utilization, and therefore losses in aggregate output. In turn, dispersion in the marginal product of firms can be rationalized using alternative mechanisms, such as the one presented above, where capital irreversibility ($v_{i,t}>0$), individual investment inefficiency ($\eta_{i,t}>1$) and firm specific collateral constraints $(\varphi_{i,t})$ can generate discrepancies between the return to capital and the real interest rate\footnote{This is similar to \cite{10.1093/qje/qjx024} who assume instead capital adjustment costs and size dependent borrowing constraints. The latter was shown to be important for capital misallocation with the decline of the real interest rate faced by the European South.}. The paper will allow for a general dependence of this dispersion on all shocks, including the real interest rate as in GKKV.
\subsubsection{\textit{Aggregate Labor Supply and Wage Determination}} Aggregating the intratemporal condition, we get that aggregate labor supply is distorted by taxes and household heterogeneity, $\Xi^{lab}_{t}$: \[L^{s}_{t}=\left(\frac{W_{t}(1-\tau^{inc}_{t})}{P_{t}}\right)^{\frac{1}{\eta}}(C_{t}-hC_{t-1})^{-\frac{\omega}{\eta}}\Xi^{lab}_{t}\] where $\Xi^{lab}_{t}=1+\frac{\omega(\omega+\eta)}{2\eta^{2}}(C_{t}-hC_{t-1})^{-2}Var(c_{i,t})$. 

Correspondingly, following \citet{GaliWAGERIG} real wage rigidity is parsimoniously modeled as follows:
\[\frac{W_{t}}{P_{t}}= \left(\frac{W_{t-1}}{P_{t-1}}\right)^{\rho_{w}}\left(MRS_{t}\right)^{1-\rho_{w}}\] where $MRS_{t}:= {L_{t}^{\eta}}/{\left(C_{t}-hC_{t-1}\right)^{-{\omega}}(\Xi^{lab}_{t})^{\eta}(1-\tau^{inc}_{t})}$. When $\rho_{w}\to 0$, real wages are competitively determined.
\subsubsection{\textit{Macroeconomic Identity}}
Aggregating the private sector budget constraint using $\Lambda_{t}(i)$ and combining it with the government budget constraint, we get that \vspace{-0.1 in}
\[I_{t}+C_{t} + G_{t} + R^{\star}_{t}D_{t}-D_{t+1} +\frac{\psi}{2}\int u^{2}_{i,t}k_{i,t}d\Lambda_{t,t}=Y_{t}\]
Furthermore, using that international bonds have to be in zero net supply, \[D^{for}_{t}+D_{t}=0\] for all $t$. Defining net foreign lending as $NFL_{t}:=D^{for}_{t+1}-R^{\star}_{t}D^{for}_{t}$, the macroeconomic identity becomes as follows:
\[I_{t}+C_{t} + G_{t} + NFL_{t} +\frac{1}{2}\alpha \frac{\nu^{p}_{t}-1}{\nu^{p}_{t}}Y_{t} =Y_{t}\]  \subsection{Informing $\mu^{c}_{t}$ and $\lambda_{I,t}$}
As in the KS model, we can proceed with estimation of the aggregate model above by using the smallest set of macroeconomic data and cross-sectional moments, and we can in principle identify how  $\mu^{c}_{t}$ and $\lambda_{I,t}$ interact with aggregate shocks. Nevertheless, one can further decompose these two distortions into an intensive and an extensive margin. 
 That is, we can decompose these distortions into the proportion of households and firms whose behavior is distorted and by how much they adjust, on average:
	  \begin{eqnarray*}
		\mu^{c}_{t}& \equiv & \mathbb{E}_{t}(\mu_{i,t}|\mu_{i,t}\neq 0)\mathbb{P}_{t}(\mu_{i,t}\neq 0)\\
			\lambda_{I,t}& \equiv & \mathbb{E}_{t}(\lambda_{I,i,t}|\lambda_{I,i,t}\neq 0)\mathbb{P}_{t}(\lambda_{I,i,t}\neq 0)
	\end{eqnarray*}
The extensive margin captures how aggregate shocks push a measure of consumers and firms against their constraints while the intensive margin captures by how much they distort their decisions on average and how  this depends on aggregate shocks. To see how these statistics relate to interesting quantities, suppose that we are interested in computing the consumption response to fiscal stimulus, financed by lump sum taxes (negative transfers), $G_{t}=\int t_{i}d\Lambda_{i}(t)$. 

Abstracting from labor supply ($l_{i,t}=1$), setting marginal tax rates equal to zero and assuming no cost of capital capital utilization ($\psi=0$) and perfect competition, cash on hand is equal to ${y}_{i,t} = W_{t} -t_{i} $. For the sake of illustration, assuming that the resulting density of the cross sectional distribution of consumption\footnote{Consumption is a function of the state variables, thus the distribution of consumption is a tranformation of the joint distribution of the underlying states.} is exponential,  $f_{t}(i)= \rho e^{-\rho c_{i,t}}$ ($\rho$ is the rate parameter, where $C_{t}=\frac{1}{\rho}$,$Var(c_{i,t})=\frac{1}{\rho^2}$), the response of aggregate consumption to an increase in government purchases is equal to:
\begin{eqnarray*}
\frac{dC_{t}}{dG_{t}}&=&\int \frac{dc_{i,t}}{dG_{t}} d\Lambda_{i}(t) +  \int c_{i,t} \frac{d(d\Lambda_{t}(i))}{dG_{t}}\end{eqnarray*} which can be analyzed as follows \begin{eqnarray*}
&&\int \frac{dc_{i,t}}{dy_{i,t}}\frac{dy_{i,t}}{dG_{t}} d\Lambda_{t}(i) +  \int c_{i,t} \frac{dln(f_{i}(t))}{dG_{t}}d\Lambda_{t}(i)\\
&=& \frac{dW_{t}}{dG_{t}}\int\frac{dc_{i,t}}{dy_{i,t}}d\Lambda_{t}(i) +  \frac{1}{2\rho}\frac{d ln(Var_{t}(c_{i,t}))}{dG_{t}}-\int\frac{c_{i,t}}{C_{t}}\frac{dc_{i,t}}{dG_{t}}d\Lambda_{t}(i)\\
&\approxeq& \frac{dW_{t}}{dG_{t}}\int\frac{dc_{i,t}}{dy_{i,t}}d\Lambda_{t}(i) +  \frac{1}{2\rho}\frac{d ln(Var_{t}(c_{i,t}))}{dG_{t}}
\end{eqnarray*} 

where we ignore the last higher order term in the last equation. Distributional effects of changes in government spending are captured by changes in the variance, just as in our case. Moreover, the average adjustment is determined by the general equilibrium response of the wage rate and the average marginal propensity to consume (MPC). 
The latter is the average of the MPC of those agents who behave as permanent income consumers (low MPC) e.g. those who are well insured or do not face any kind of constraint, and those agents with high MPC e.g. the constrained (poor or otherwise). 

The representative agent point of approximation captures permanent income type of behavior while the aggregate response of the constrained is:
   \[\int_{\mu_{i,t}\neq0}\frac{dc_{i,t}}{dy_{i,t}}d\Lambda_{t}(i)=\underset{Intensive}{\underbrace{\int \frac{dc_{i,t}}{dy_{i,t}}f_{t}\left(i: \mu_{i,t}\neq0\right)di}}\underset{Extensive}{\underbrace{\int_{\mu_{i,t}\neq0} f_{t}(i)di}}\]
These are two aggregate variables that depend on aggregate states and this dependence could be potentially separately identified, if one can measure either of them at business cycle frequencies. 
This paper's approach identifies the corresponding log-linear relation between the margins of adjustment and the shocks\footnote{One could of course find other ways or other moments to meaningfully inform these distortions. Direct estimates of the marginal propensities to consume of household groups at the business cycle frequency could also be employed, but this entails its own measurement issues, as well as the difficulty of obtaining  reliable measures at this frequency.}. Moreover, had we allowed for labor choice in this example, we would have to identify how the covariance of the marginal propensity to consume and the marginal propensity to work vary with aggregate shocks.
\section{Taking the model to the data}Recall that the representation result in Proposition \eqref{repre} in Section 4 implies that there is a one to one mapping from distortions in the Euler equation to distortions in the law of motion, and these distortions will depend on the endogenous and exogenous aggregate states. The Euler equation is therefore augmented with these distortions. For maximal generality, the variance of the consumption shares is allowed to depend on all shocks and the capital stock\footnote{Recall that from Section 3,  $Var_{t}(c_{i,t+1})$ is computed under $p(s_{i,t+1}|S_{t+1},S_{t}) ds_{i,t+1}$ and it therefore depends potentially on all aggregate states. Observing measures of consumption variance over time, permits identification of this channel without resorting to a specific model.}. Solving the model using conventional techniques allows us to trace the effect of both the distortions and cross sectional moments (heterogeneity) to all macroeconomic variables. The log-linearized estimable model is therefore as follows, where $\Lambda_{int}$ and $\Lambda_{ext}$ denote the coefficients loading on the states and correspond to the elasticities of the intensive and extensive margins to aggregate shocks for both consumption  and investment: \vspace{-0.4 in}  \begin{spacing}{1.5}     \small
 \begin{eqnarray}
  \nonumber 	\tilde{C}_{t}&=&  h\tilde{C}_{t-1} + \beta R_{ss}\left( {\mathbb{E}_{t}}(\tilde{C}_{t+1}-h\tilde{C}_{t})
  -\frac{1-h}{\omega}{\mathbb{E}_{t}}\left(\tilde{R}_{t+1}+(\tilde{\Xi}_{t+1}-\tilde{\Xi}_{t})\right)\right)\\
  \nonumber &&-(1-\beta R_{ss})\frac{1-h}{\omega}\mu_{c,t} \\
 \nonumber	\mu_{c,t}&=& \Lambda_{int}^{c'}\tilde{S}^{c}_{t} + \tilde{B}^{con}_{t} \equiv \Lambda_{int}^{c'}\tilde{S}^{c}_{t}+ \Lambda_{ext}^{c'}\tilde{S}_{t}\\
 \nonumber	\tilde{W}_{t}&=& \rho_{w}\tilde{W}_{t-1} +(1-\rho_{w})\left(\frac{\omega}{1-h}\left( \tilde{C}_{t}-h\tilde{C}_{t-1}\right)+\eta\tilde{L}_{t}+\frac{\tau^{inc}_{ss}}{1-\tau^{inc}_{ss}}\tilde{T}^{inc}_{t}-\tilde{\Xi}^{lab}_{t}\right)\\
 		\nonumber	\tilde{W}_{t} &=& \tilde{Y}_{t} - \tilde{L}_{t} - \tilde{\nu}^{p}_{t} \\
\nonumber	\tilde{K}_{t+1}&=&(1-\delta )\tilde{K}_{t}+(\delta-\lambda_{I,ss}) \tilde{I}_{t} + \lambda_{I,ss}\lambda_{I,t}\\
\nonumber	\lambda_{I,t}&=& \Lambda_{int}^{I'}\tilde{S_{t}} + \tilde{B}^{inv}_{t}\\
\nonumber	\tilde{B}^{inv}_{t} &=& \Lambda_{ext}^{I'}\tilde{S_{t}}\\
\nonumber	 \tilde{K}^{u}_{t} &= &  \tilde{K}_{t-1}+ \hat{U}_{t}=  \tilde{K}_{t-1}+ \frac{1}{2}\left(\tilde{r}^{k}_{t}+ \frac{1}{v^{p}_{ss}-1} \tilde{\nu}^{p}_{t}\right)+\frac{{v}^{k}_{ss}}{{v}^{k}_{ss}-8} \tilde{V}^{k}_{t}\\
\nonumber	\tilde{Y}_{t} &=& \alpha\tilde{Z}_{t}+\alpha \tilde{K}^{u}_{t}+(1-\alpha)\tilde{L}_{t} \\
\nonumber	\tilde{r^{k}_{t}} &=& \alpha\tilde{Z}_{t}-(1-\alpha )\tilde{K}_{t} +(1-\alpha)\tilde{L}_{t}\\
\nonumber	y_{ss}\tilde{Y_{t}} &=&  \left(1-\frac{\alpha}{2}\frac{v^{p}_{ss}-1}{v^{p}_{ss}}\right)^{-1}\left(c_{ss}\tilde{C}_{t}+i_{ss}\tilde{I}_{t} + g_{ss}\tilde{G}_{t}  + nfl_{ss}\widetilde{NFL}_{t}+ \frac{\alpha}{2} \frac{y_{ss}}{v^{p}_{ss}}\tilde{\nu}^{p}_{t} \right)\\
 \widetilde{V}^{k}_{t} &= &  vk_{p}\tilde{B}^{inv}_{t}+vk_{z}\tilde{Z}_{t} + vk_{for}\widetilde{NFL}_{t} + vk_{k}\tilde{K}_{t}  + vk_{g}\tilde{G}_{t}  + vk_{r}\tilde{R}_{t}  + vk_{\tau}\tilde{T}^{lump}_{t} \label{40} \\
\nonumber   \tilde{R}_{t} &= &  \rho_{R}\tilde{R}_{t-1} +\epsilon_{R,t} \\
 \nonumber  Tr_{ss} \tilde{Tr}_{t} &= &  \tau^{inc}_{ss}W_{ss}L_{ss}(\tilde{T}^{inc}_{t}+\tilde{W}_{t}+\tilde{L}_{t}) + \tau^{corp}_{ss}Pr_{ss}(\tilde{Pr}_{t}+\tilde{T}^{corp}_{t})+\tilde{T}^{lump}_{t}  \\
\nonumber     &= & \frac{1}{1+\phi_{d}}\left(\phi_{d}g_{ss}\tilde{G}_{t}+(\phi_{d}-\phi_{f})nfl_{ss}\widetilde{NFL}_{t}+\epsilon_{Tr,t}\right) \notag\\
\nonumber         \tilde{\tau}^{inc}_{t} &=& \tilde{\tau}^{corp}_{t} + \epsilon_{T,t} \\
\nonumber     \tilde{Pr_{t}} &= &\tilde{Y_{t}}+\frac{1}{\nu^{p}_{ss}-1}\tilde{\nu}^{p}_{t}\\
  \widetilde{V}^{c}_{t} &= &  vc_{p}\tilde{B}^{con}_{t}+vc_{z}\tilde{Z}_{t} + vc_{for}\widetilde{NFL}_{t} + vc_{k}\tilde{K}_{t}  + vc_{g}\tilde{G}_{t}  + vc_{r}\tilde{R}_{t}  + vc_{\tau}\tilde{T}^{lump} \label{variancec} \\
 \tilde{\Xi}_{t}&=&\frac{\omega(1+\omega)V^{c}_{ss}}{2(1-h)^{2}+\omega(1+\omega)V^{c}_{ss}}\left(\widetilde{V}^{c}_{t}-\frac{2h}{1-h}(\tilde{C}_{t}-\tilde{C}_{t-1})\right) \nonumber\\
 \tilde{\Xi}^{lab}_{t}&=&\frac{\omega(\eta+\omega)V^{c}_{ss}}{2\eta^{2}(1-h)^{2}+\omega(\eta+\omega)V^{c}_{ss}}\left(\widetilde{V}^{c}_{t}-\frac{2h}{1-h}(\tilde{C}_{t}-\tilde{C}_{t-1})\right)\nonumber
 \end{eqnarray} \end{spacing}\normalsize 
where $\tilde{S}_{t}=(\tilde{Z}_{t},\widetilde{NFL}_{t},\tilde{K}_{t-1},\tilde{G}_{t},\tilde{R}_{t},\tilde{T}^{lump})$ are the aggregate states that are considered to be the potential drivers of financial frictions.

 When accounting for intensive distortions, I also control for heterogeneity within the intensive margin to separately identify possible dependence on aggregate shocks that feeds through heterogeneity in individual consumption and corporate income. I thus let $\tilde{S}^{c}_{t}=(\tilde{S}_{t}, \widetilde{V}^{c}_{t},  \widetilde{V}^{k}_{t})$. 
  

Besides aggregate states, to improve the identification of the variance $\widetilde{V}^{c}_{t}$ when it is latent, the latter is explicitly linked to the proportion of constrained agents $\tilde{B}^{con}_{t}$ using the law of total variance:
\small
\begin{eqnarray}
{V}^{c}_{t}&=& B^{con}_{t}\left( \mathbb{E}_{t}\left[\left(\frac{c_{i,t}}{C_{t}}\right)^2 \mid \lambda_{i,t}>0\right]-\mathbb{E}_{t}\left[\left(\frac{c_{i,t}}{C_{t}}\right)^2 \mid \lambda_{i,t}=0\right] \right)\nonumber\\
&& + \mathbb{E}_{t}\left[\left(\frac{c_{i,t}}{C_{t}}\right)^2 \mid \lambda_{i,t}=0\right]-1 \label{eq:cons_var}
\end{eqnarray}\normalsize
The first component is the difference between the second moments of constrained and unconstrained agents while the second component is a measure of variation across unconstrained agents. If nobody is constrained, then there is no "between" variation. Log-linearizing this expression leads to \eqref{variancec}, where it is recognized that the "between" variation and the variance of the unconstrained cannot be separately identified but can be linked to all the shocks. 

Finally, for $\epsilon\sim N(0,diag(\Sigma))$, the exogenous variables evolve as follows : \vspace{-0.3 in} 
\begin{spacing}{1.2}
\begin{eqnarray*}
	\tilde{Z}_{t} &= &  \rho_{z}\tilde{Z}_{t-1}+ \epsilon_{Z,t}\\
	\widetilde{NFL}_{t} &= &  \rho_{nfl}\widetilde{NFL}_{t-1}+ \epsilon_{NFL,t}\\ 
	\tilde{G}_{t} &= &  \rho_{G}\tilde{g}_{t-1}+ \epsilon_{G,t} + c_{gz}\epsilon_{,t}\\
	\tilde{R}_{t} &= &  \rho_{r}\tilde{R}_{t-1}+ \epsilon_{R,t}\\
	\tilde{T}^{lump}_{t} &= &  \rho_{\tau}\tilde{T}^{lump}_{t-1}+ \epsilon_{T^{lump},t}\\
	{v}^{p}_{t} &= &  \rho_{vp}{v}^{p}_{t-1}+ \epsilon_{v^{p},t} + \rho_{vp,g}\epsilon_{G,t}+\rho_{vp,r}{v}^{p}_{t-1}
\end{eqnarray*} \end{spacing}  \normalsize \vspace{-0.1 in} 
Due to the small open economy assumption, the real interest rate is taken as exogenous. Moreover, despite that there are no micro-foundations, the inclusion of a time varying price markup absorbs any wage variation that cannot be explained by average productivity, including wage rigidities, and captures the effects of policy shocks on markups without resorting to a particular mechanism of price rigidity. This is important for avoiding misspecification in the construction of counterfactuals\footnote{In the Appendix I plot standardized estimates of the (inverse) markup: It does not explain much of the wage during the crisis, while they broadly co-move before 2007. Correlation s statistically insignificant ((-0.34, 0.13)).} as well as avoiding taking a stance on how interest rate and fiscal policy shocks affect firm profitability\footnote{The benchmark New Keynesian model has been criticized for its implications regarding the cyclicality of the markup when prices are rigid, see e.g. \citep{RePEc:nbr:nberch:14366, 10.1093/restud/rdy060} and references therein. }.  

\subsection{Data, Trends and Measurements} The model is estimated using time series data from 1999Q1 to 2018Q1 on: aggregate (durable and non-durable) consumption, hours, gross capital formation, capacity utilization, government spending, the ex-post real EONIA rate, income and corporate tax rates, tax revenues as $\%$ of nominal GDP, the estimates of GKKS for the dispersion of the marginal product of capital across Spanish firms\footnote{I thank Loukas Karabarbounis for providing me with this measure.}, two measures of cross sectional dispersion of consumption obtained from the Survey of Household Finances and the Household Budget survey respectively and two measures on financially constrained consumers and firms, $(B^{con}_{t},{B}^{inv}_{t})$ respectively. While FCI can be considered as a proxy for the proportion of financially constrained firms\footnote{In the model borrowing constraints affect firms primarily through investment, and this is in accordance with the evidence presented in Figure \ref{spain}. I thus employ the FCI series as a proxy and not FCP. Please consult the Appendix for the exact wording and transformations.}, the proportion of borrowing constrained consumers is a latent variable that is harder to measure. 

The FBC measure plotted in Figure \ref{spain} is the proportion of households that responded that they are "running into debt" in the Business and Consumer Survey (BCS) of the European Commission\footnote{This comprises of monthly data, which can be downloaded from \href{https://ec.europa.eu/info/business-economy-euro/indicators-statistics/economic-databases/business-and-consumer-surveys_en}{here}.}. This is an  an upper bound to the true fraction of households that are likely to face borrowing constraints within the quarter, as some of those already into or getting into debt  may (have been) be denied a loan application. 

As in \citet{Tryphonides2020}, I combine 
FBC with a direct measure computed from the Survey of Household Finances (SHF). This measure asks households whether they have been denied a loan application during this period\footnote{Accounting for consumers that have not applied for loans because they expect to be rejected (also coined as "discouraged borrowers" in \citet{10.2307/2077836} ) or consumers that were granted a fraction of the requested loan does not affect the estimate of this proportion.}. In fact, such questions also ask for the reason for which these households have been rejected, and these reasons vary depending on individual characteristics, employment, guarantees, changes in the institution's credit policy, excessive debt etcetera. Yet, this measure is available every three years.

\begin{figure}[H]
	\begin{centering}			
		\includegraphics[scale=0.5]{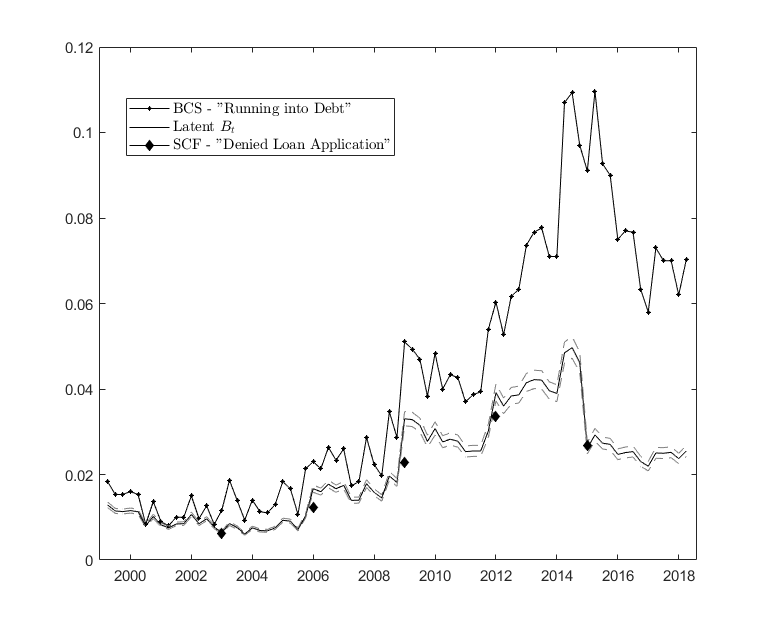}
		\par
	\end{centering}  
	\protect\caption{}
	\label{mixedLCC}
\end{figure}

 To infer the true proportion of constrained households for the periods in which it is not observed, so all periods between the triennial measure from SHF, I employ a mixed frequency model, where the quarterly observations on consumers running into debt are linked to the triennial exact measure.  
 Letting  $B^{con}_{t}$ be the exact measure and $\Pi_{t}$ the proportion of consumers running into debt, then $B^{con}_{t}=\zeta_{t}\Pi_{t}$ where $\zeta_{t}\in(0,1)$, as the agents that hit the minimum level of permissible debt are a fraction of the total measure of indebted households. As shown in the appendix, this serves as a measurement equation in a mixed frequency state space model. I plot the extracted measure in Figure \ref{mixedLCC}.

Furthermore, excluding durables from aggregate consumption does not seem a reasonable thing to do since  liquidity constraints are more binding in the case of large purchases e.g. automobiles. As far as housing is concerned, consumption measures include imputed rents, which can be thought as the service value of housing and not the housing stock per se. 

Consistent with the choice of observable, housing choice is not separately modeled\footnote{As an approximation, accumulating capital stock that produces a level of income after paying a cost of labor is equivalent to accumulating housing stock that, with a certain level of labor input such as housework produces a service flow that enters consumption.}. 
Moreover, to compute the variance of consumption in the two surveys durable consumption is included as well. More details on the exact computation are in  the Appendix.

\subsubsection{Measurement Equations for Cross Sectional Variances}
Cross section quantitative surveys are available on an annual basis and it is assumed that reported annual consumption is the sum of quarterly consumption (up to measurement error). Also, consistent with the partial equilibrium example in \citet{Tryphonides2020}, the reduced form process for consumption under liquidity constraints is an autoregressive process. Therefore, denoting by $\rho_{c}$ the persistence component of the individual consumption share $\hat{c}_{i,t}:=\frac{c_{i,t}}{C_{t}}$:
\begin{eqnarray*}
	Var(\hat{c}^{ann}_{i,t}) 
	&=&  Var(\hat{c}^{Q4}_{i,t})+ (1+2\rho_{c})Var(\hat{c}^{Q3}_{i,t})+\\
	&&  +(1+2\rho_{c}(1+\rho_{c}))Var(\hat{c}^{Q2}_{i,t}) + +(1+2\rho_{c}(1+\rho_{c}+\rho^{2}_{c}))Var(\hat{c}^{Q1}_{i,t})
\end{eqnarray*}\normalsize

The same measurement equation is applied to the variance of the marginal product of firms, where $mpk_{i,t}$ is also assumed to be an autoregressive process of order one with coefficient $\rho_{mpk}$. 

\subsubsection{Trends}
The presence of frictions, aggregate shocks and a slowly evolving distribution of consumers and firms makes it difficult to argue that the model does not have implications for frequencies lower than the typical business cycle frequency. A casual inspection of the macroeconomic and cross sectional moment time series verifies this possibility. At the same time, the model is approximated around a non-stochastic steady state, implying that when appropriately scaled, all observables should have a meaningful long run average or in other words, they should fluctuate around a trend that possibly depends on the permanent component of shocks. Therefore, removing variation in the observables without taking into account their joint determination is likely to bias our results. Following  \citet{CANOVA20141}, I embed the model in a more general measurement system akin to a local level model, where I allow for a common permanent component to consumption and its dispersion, investment, wages and government spending. More particularly, in the notation of Section 4, the solution $(P^{\star}(\theta),Q^{\star}(\theta))$ of the model is embedded to a more general system where the above mentioned observables are modelled as a sum of a permanent trend and a non-permanent component, where the permanent component is a random walk and $X^{np}_{t}$ is the law of motion of the model:
\begin{eqnarray*}
  X^{obs}_{t} & = & S_{\Upsilon}\Upsilon_{t} +  S_{X}X^{np}_{t}\\
 \Upsilon_{t} & = & \Upsilon_{t-1} +  \epsilon^{p}_{t}\\
   X^{np}_{t}& = &P^{\star}(\theta) X^{np}_{t-1} + Q^{\star}(\theta)\epsilon^{np}_{t}
\end{eqnarray*} 

where $S_{\Upsilon}$ and $S_{X}$ are the linking functions from measurements to states\footnote{For more details on $S_{\Upsilon}$ and $S_{X}$ please see the Appendix.}. Furthermore, I allow for $5\%$ of the variance of all aggregates is measurement error. 

\subsubsection{Prior distributions and Estimation} I assume largely uninformative priors, which are presented in Table \ref{table:ParamPriors}
 in the Appendix. A 2-block RW-MCMC is used to obtain posterior parameter draws,  while  $\Theta^{CS}_{5\%}$ is constructed using (projections) of the $95\%$ joint confidence set obtained using quantiles from the posterior draws of the objective function as in \cite{doi:10.3982/ECTA14525} in order to take into account potential failures of point identification.  
 The confidence sets are reported in Table \ref{table:ParamEst}.

\subsection{Results}
Parameter estimates yield several insights both for the effects of aggregate shocks on the corresponding margins. Productivity, government spending and foreign lending shocks are quite persistent while tax shocks are less persistent but with higher conditional volatility than the rest of the shocks. Productivity shocks have a positive effect on government spending, suggesting that spending is pro-cyclical. 
Moreover, the weight in the tax revenue target rule on domestic debt is higher than foreign debt, probably driven by the increased holdings of government debt by domestic banks (2008-2012) and the subsequence increase in holdings by the Bank of Spain, as a result of the Public Sector Purchase Programme and its conditionalities. 
Regarding markups, what is interesting is that there is no posterior evidence for the cyclicality with respect to either productivity shocks or real interest rate shocks, while there is evidence that the markups are countercyclical with respect to government spending. 

\begin{flushleft}	
	\begin{table}
			\caption{Robust Confidence Set for $\Theta$}\vspace{-0.1 in}
			\label{table:ParamEst}
			\begin{tabular}{c|c|c|c|c|c|c|c|c|c|}
				\hline
				$\Theta$ & $q_{5\%}$ & $q_{95\%}$ & $\Theta$ & $q_{5\%}$ & $q_{95\%}$  & $\Theta$ & $q_{5\%}$ & $q_{95\%}$ \\ 
				\hline
				\multicolumn{6}{c|}{Preferences} & \multicolumn{3}{c|}{Cons. Heterogeneity} \\  
				$\omega$ &  3.885 & 3.999&   $\eta$ &6.418 & 7.226& $V^{c}_{ss}$ & 0.01371 & 0.0166\\				
				$ \beta$& 0.9819 & 0.9839  & 	$h$ &0.2138 & 0.4102& $vc_{p}$ &-0.2204 & -0.1037\vspace{0.05 in}\\  
				&\multicolumn{2}{c|}{Exogenous Shocks} &\multicolumn{3}{c|}{Debt Policy} & 	$vc_{z}$ &-3.802 & -2.244\\ 
				$\rho_{z} $& 0.863 & 0.9446 &  	$\phi_{d}$ & 24.660 & 27.950 & 	$vc_{g}$ & 0.1456 & 0.9943\\
				$\rho_{g}$ & 0.9277 & 0.9701&		    $\phi_{f} $&7.043 & 9.695& $vc_{nfl}$ & -2.789 & -2.319 \\
				$\rho_{nfl}$&  0.9606 & 0.9772 &	$\sigma_{\epsilon_{Tr}}$ &0.0658 & 0.1122& $vc_{k}$ &1.605 & 2.570 \\				
				$\rho_{r,1}$&0.8130 & 0.9240&	\multicolumn{3}{c|}{Price Dispersion} &$vc_{r}$ &1.374 & 2.344\vspace{0.05 in}\\  	 
				$\rho_{r,2}$&0.0707 & 0.1141 & $pd_{ss}$ &0.0741 & 0.8810 & $ vc_{\tau}$ &  -0.3590 & 0.5595\\
				$\rho_{\tau}$ &0.6116 & 0.8689 & \multicolumn{3}{c|}{Rigidity and Markup}      &$v_{c,id}$ &1.376 & 2.007	\\
				$c_{gz}$&0.3047 & 0.652 &	$\rho_v^{p}$ &0.9452 & 0.9626& \multicolumn{3}{c|}{Firm Heterogeneity}\vspace{0.05 in}\\ 				
				$\sigma_z$ & 0.0520 & 0.0682& 		 $\sigma_{v^{p}}$ &0.0398 & 0.0475 & $vi_{p}$ &-11.36 & -9.582\\
				$\sigma_g$&0.0401 & 0.0485  &	 $\rho_{v^{p},g}$ &-0.7430 & -0.1990& $vi_{z}$ & 4.118 & 4.431 \\			
			    $\sigma_{nfl}$ &0.0669 & 0.0970  & $\rho_{v^{p},m}$ &-1.753 & 0.4145& $vi_{g}$&-2.344 & -0.6431\\	 
			  	$\sigma_{r}$ &0.0073 & 0.0090& $\rho_{v^{p},z}$ &-0.1334 & 0.1893& $vi_{nfl}$&1.341 & 5.594\\ 
				$\sigma_{\tau}$ &  0.3427 & 0.6457& $\rho_{w}$ &0.7385 & 0.8338 & $vi_{k}$&-0.6559 & 0.9013\\ 					
				\multicolumn{3}{c|}{Constrained Consumers}     & 	\multicolumn{3}{c|}{Constrained Inv.  Firms} &$vi_{\tau}$&-6.451 & -4.367  \\ 				
			   $\lambda^{c,ext}_{z}$ & -2.881 & -1.683&   $\lambda^{I,ext}_{z}$ &-1.056 & -0.2335 &$vi_{r}$& -2.778 & -2.264 \\
			 $\lambda^{c,ext}_{g}$ &0.5816 & 2.53 &	$\lambda^{I,ext}_{g}$ &1.901 & 2.332             &     $v_{i,id}$ & 0.0670 & 0.2905             \\
			$\lambda^{c,ext}_{nfl}$ & 0.0747 & 0.9848	& $\lambda^{I,ext}_{nfl}$&-1.0590 & -0.7936\\ 
				$ \lambda^{c,ext}_{k}$ &5.594 & 6.335& $\lambda^{I,ext}_{k}$ & 1.027 & 2.400 & \multicolumn{3}{c|}{Measurement-Dispersion}\\ 				
			  	$\lambda^{c,ext}_{\tau}$ & 0.1506 & 0.327& $\lambda^{I,ext}_{\tau} $&-0.6784 & -0.4121 &$\rho_{mpk}$  &-0.999 & -0.9625 \\  	  	
			$\lambda^{c,ext}_{r}$ &17.69 & 20.07&$\lambda^{I,ext}_{r}$&11.2 & 16.15& $\rho_{c}$ &  -0.999 & -0.9672\\ 				
				$\lambda^{c}_{z}$&1.298 & 3.005&	$\lambda^{I}_{z}$ &-1.622 & -1.282\\ 				
				$\lambda^{c}_{g}$ &5.123 & 8.139 &	$\lambda^{I}_{g} $&-3.292 & -3.107\\			
			$\lambda^{c}_{nfl}$&7.838 & 9.971&	$\lambda^{I}_{nfl}$ & 0.6974 & 2.85\\						
		    $\lambda^{c}_{k} $&-6.596 & -4.704&  	$ \lambda^{I}_{k}$& -0.3325 & 0.6473\\ 	 			
		$\lambda^{c}_{\tau}$& -2.993 & 1.731&$\lambda^{I}_{\tau}$& -0.1325 & 1.18\\  	 			
				$\lambda^{c}_{r}$ & 2.713 & 4.727&$ \lambda^{I}_{r}$ & -2.462 & -1.771 \\			   
				$\lambda^{c}_{V^{k}}$&0.4435 & 0.6007 &	$\lambda_{I,ss}$ & -0.0020 & -0.0003\\ 				
					$\lambda^{c}_{V^{c}}$ &6.233 & 6.738\\				
						  \end{tabular}
				\end{table}
\end{flushleft}

\subsubsection{Conditional Consumption Behavior}
In terms of relative magnitudes, exogenous shocks in the ex-ante real rate shocks are the most important drivers of the extensive margin in consumption. An increase in the real rate leads to a $\approx 19\%$ increase in the proportion of the borrowing constrained. While a higher interest rate encourages savings, those households that are still willing to borrow cannot service the higher cost of borrowing. The effect on the  intensive margin is  also positive  but  smaller ($\approx 3.5\%$). A higher rate increases the returns to bond holders but it also increases the debt interest payments on existing debt.  A second round effect arises as consumption dispersion increases by $\approx 1.8\%$ due to the direct interest rate effect, an expected outcome given the heterogeneity of debt levels and the differential effects on savers versus borrowers. Moreover, the initial rise in the proportion of the borrowing constrained leads to a small dampening effect on consumption dispersion. Inspecting \eqref{eq:cons_var}, the negative coefficient $vc_{p}$ suggests that the dispersion in consumption by the unconstrained consumers wealky dominates that of the constrained in the steady state.

On the other hand, productivity shocks decrease the share of borrowing constrained by $\approx 2.5\%$ as they increase employment and wages, while they increase the distortion to the intensive margin by roughly the same percentage. 
Consumption dispersion falls, which leads to a fall in aggregate labour supply. A second round dampening effect arises as the decrease in the proportion of constrained consumers increases consumption dispersion. Government spending shocks seem to affect the consumption margins differently. They contribute to a  $\approx 1.5\%$ increase in the share of the borrowing constrained and a $ \approx 6.5\%$ on the intensive margin. Both can have a negative impact on aggregate consumption. The increase in government spending increases dispersion as well, while the direct effect is also dampened by the increase in the extensive margin. Lump sum tax shocks do not seem to contribute much in the consumption margins, appart from a slight increase in the proportion of the borrowing constrained. Net foreign demand shocks, or negative foreign borrowing shocks, affect both the intensive and the extensive margin of the  borrowing constrained. This suggests that exogenous constraints to international borrowing also exert distortions to domestic consumption.

\subsubsection{Conditional Investment Behavior}
On the investment side, shocks to the ex-ante real interest rate are also (conditionally ) important for the behavior of the extensive margin as a $1\%$ increase in the rate increases the proportion of investment inefficient firms by $\approx 13.5\%$. An increase in the cost of funding necessarily causes certain firms to become investment constrained, while on average, the effect on the intensive margin is negative. 
On the other hand, productivity shocks seem to lower both the extensive and the intensive margin of the distortion. Foreign demand shocks lower the extensive margin while they increase the intensive margin of adjustment.  

In general,  high dispersion in the marginal product of capital ($mpk_{i}$) leads to lower utilization of the existing capital stock, leading to output - and therefore welfare - losses. Moreover, there is evidence that dispersion is very much dominated by that of the currently unconstrained firms $(vi_{p}<0)$. What this suggests is that to some extent, it is the existence of e.g. capital adjustment costs, which are also a source of heterogeneity in the marginal product of capital that can explain the latter rather than occasionally binding constraints. Productivity shocks increase dispersion both directly and indirectly through increases in $\tilde{B}_{t}^{inv}$. Same holds for government spending shocks, which decrease dispersion both ways, suggesting that spending might alleviate constraints and inefficiencies. This is also consistent with the fact that while government spending seems to crowd out investment by increasing the proportion of firms that face an increasing cost of funding, for those that have been constrained before the spending shock, it may lower the degree by which they distort their investment choice.

In terms of unconditional explanatory behavior, the variance decomposition in Table \ref{table:VarDec} illustrates that
shocks to government spending and net foreign borrowing seem to explain around $30-40\%$ of the variance of the endogenous variables (with the exception of tax revenues, where foreign lending shocks explain $58\%$), while productivity shocks explain around    $10-15\%$ of all variables except consumption, where they explain a share of $24\%$. So while the ex-ante reall rate can explain a lot of the conditional variation, this is not true in an unconditional sense. This can be explained by the relatively lower variance of the shock across the sample. Posterior evidence also points towards less idiosyncratic risk in investment than consumption.

\vspace{-0.2 in}

\footnotesize
\begin{flushleft}	
	\begin{table}[H]
		\caption{Uconditional Variance Decomposition, $\%$ (at mode of $\Theta$)}\vspace{0.1 in}
		\label{table:VarDec}
		\begin{tabular}{c|c|c|c|c|c|c|c|c|c|c|} 
	&$\epsilon_{Z,t}$ & $\epsilon_{G,t}$ & $\epsilon_{NFL,t}$ & $\epsilon_{R,t}$ &$\epsilon_{\tau,t}$& $\epsilon_{T,t}$ & $\epsilon_{v^{p},t}$ & $\epsilon_{Tr,t}$ & $\epsilon_{V_{c},t}$ & $\epsilon_{V_{I},t}$\\ \hline
$C$	&23.9 & 34.5 & 26.8 & 2.5 & 0.1 & 0.9 & 11.4 & 0.0 & 0.1 & 0.0\\  
$Y$	&15.6 & 38.3 & 30.1 & 1.9 & 0.1 & 1.1 & 12.9 & 0.0 & 0.1 & 0.00\\ 
$I$	&15.4 & 38.4 & 30.1 & 1.9 & 0.1 & 1.1 & 13.0 & 0.0 & 0.1 & 0.00\\  
$H$	&15.2 & 36.6 & 28.9 & 1.9 & 0.1 & 2.3 & 15.0 & 0.0 & 0.1 & 0.00\\  
$W$	&15.4 & 37.4 & 28.3 & 2.1 & 0.1 & 0.9 & 16.0 & 0.0 & 0.1 & 0.00\\ 
$v^{p}$	 &0.0 & 24.8 &- & 2.3 &- & - & 72.9 &- &- & -\\  
$Tr$	 &11.7 & 29.6 & 58.2 & 0.00 & 0.0 & 0.0 &0.00 & 0.4 & 0.0 &0.0\\ 
$\tau^{inc}$	 &15.2 & 37.4 & 29.3 & 1.9 & 0.1 & 2.8 & 13.3 &0.0 & 0.0 & 0.0\\ 
$\tau^{corp}$	 &14.7 & 36.3 & 28.4 & 1.9 & 0.1 & 5.7 & 12.9 & 0.0 & 0.0 & 0.0\\  
$V^{c}$	 &12.1 & 26.9 & 27.9 & 1.4 & 0.1 & 0.7 & 9.1 &0 & 22.0 & 0.0\\  
$V^{k}$	 &15.1 & 38.2 & 30.3 & 2.6 & 0.1 & 1.0 & 12.7 & 0.0 & 0.0 & 0.0\\  
$B^{con}$	 &15.4 & 38.4 & 30.2 & 2.1 & 0.1& 1.0 & 13.0 & 0.0 & 0.0 & 0.0\\ 
$B^{inv}$	 &15.0 & 37.8 & 29.7 & 2.6 & 1.4 & 1.0 & 13.0 & 0.0 & 0.0 & 0.0  					  \end{tabular}	
\end{table}
\end{flushleft}   
\normalsize
\vspace{-0.3 in}

While impulse responses to all shocks could be potentially interesting, I next focus on  the government spending multiplier, a long standing empirical and theoretical question\footnote{See \citet{RAMEY201671} for a summary of estimated government multipliers in the literature.}.
%
  Although producing credible evidence on these effects is not void of challenges, several papers have tried to reconcile the identified effects from SVAR/EVAR models
   using modified versions of the neoclassical or the New-Keynesian model. On the consumption side, a notable example is \citet{RePEc:tpr:jeurec:v:5:y:2007:i:1:p:227-270} who argue that the standard New-Keynesian model can produce a positive effect of government purchases on consumption only if we account for rule of thumb consumers and price rigidities. Furthermore, on measuring the effects of fiscal policy in recessions, \citet{10.2307/23071728} reiterates the importance of controlling for the "dependence of the effects of a policy on the states of different agents, which is a key component of the dependence of the general-equilibrium effects of fiscal policy on the state of the economy".  The paper's methodology is well suited for this purpose. 
   
   Moreover, despite that the model does not specify a specific mechanism through which markups respond to changes in government spending, it accommodates for such effects. Since markups are identified through the aggregate intertemporal labor supply condition, we cannot distinguish between the conditional cyclicality of wage and price markups. Nevertheless, we can compute by how much does the response to government spending shocks changes when the contribution of markups is shut down.   
   
   Recall that the period under consideration corresponds to both years during which there was a surge in sub-national government spending e.g. 1999-2007  \citep{RePEc:bde:wpaper:1620}, while later years correspond to the face of fiscal adjustment during the crisis after 2007. The model employed in this paper features both deficit financing as well as distortionary and non-distortionary taxation, which can capture reasonably well the macro and distributional data, as well as the characteristics of fiscal policy during the period\footnote{See Figure \ref{predict} for an assessment of the fit of the model.}.  Fiscal multipliers are computed in cumulative, present value form as in e.g \citet{doi:10.1002/jae.1079}, 
 The government spending multiplier is estimated to be between $0.7$ and $1.7$, supporting therefore values well above one on impact, while negative values receive positive posterior probability after two years. These estimates correspond to the upper end of multipliers in the literature, mostly with US data, as surveyed in \cite{10.1257/jep.33.2.89}.  

\begin{figure}[H]
 	\begin{centering}
 		\includegraphics[scale=0.4]{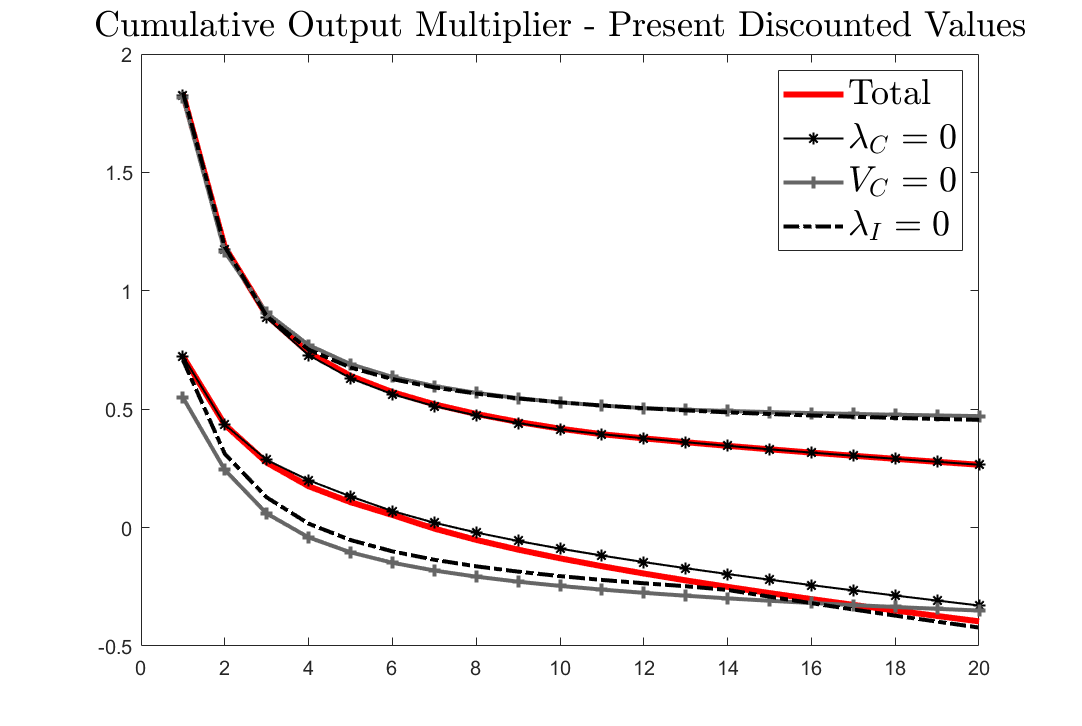} 	\vspace{-0.15 in} 	 		\par	\end{centering} \vspace{-0.15 in} 
 	\protect\caption{$90\%$ Confidence Sets.}
 	\label{out_mult}
 \end{figure}

In the next set of results I investigate several counterfactuals regarding the government spending fiscal multiplier. In particular, Figure \ref{out_mult} presents what would the multiplier be in a world in which there are no fluctuations in borrowing constrained consumers, consumption dispersion and investment constrained firms. This is an unconditional exercise, in the sense that the whole mechanism is shut down rather than just the component of the mechanism that depends on fiscal policy. The extensive margin seems to affect the multiplier in the longer term but not in a quantitatively significant way, pushing towards higher values. In the absence of investment constrained firms as well as heterogeneity in consumption, the lower bound to the multiplier is significantly reduced, supporting values as low as 0.55 on impact, as well as crossing zero as early as in 1.5 years. On the contrary, in longer horizons, shutting down these channels implies higher fiscal multipliers. 

In order to understand better the implications of the model, I next study the importance of different channels for the mutiplier by shutting down the components of different channels that depend on the government spending shock.  

Compared to the total response, the components that seem to contribute most to the final response of output by reacting to the fiscal policy shock are the markups, the intensive margin of constrained consumers, consumption and marginal revenue product dispersion, as well as the extensive margin of investment constrained firms. Markups have a positive contribution to the multiplier. Since they are estimated to be countercyclical conditional on government expenditure, an exogenous shock to the latter lowers firms profits which induces an increase in output and employment - a standard mechanism in the New Keynesian literature. 

In the absence of consumption share heterogeneity as well as the intensive margin of adjustment of constrained consumers, the response of output is much lower, while it reaches the negative territory after $\approx 1.5$ quarters. On the contrary, dispersion in the marginal revenue product of firms, a sign of inefficiency, suppresses the effect of a fiscal expansion on output at all periods. In the absence of the extensive margin of investment constrained firms, the response would much higher while positive dispersion in the marginal revenue product puts downward pressure to the multiplier across all periods.   

  \begin{figure}[H]
	\begin{centering}
		\includegraphics[scale=0.5]{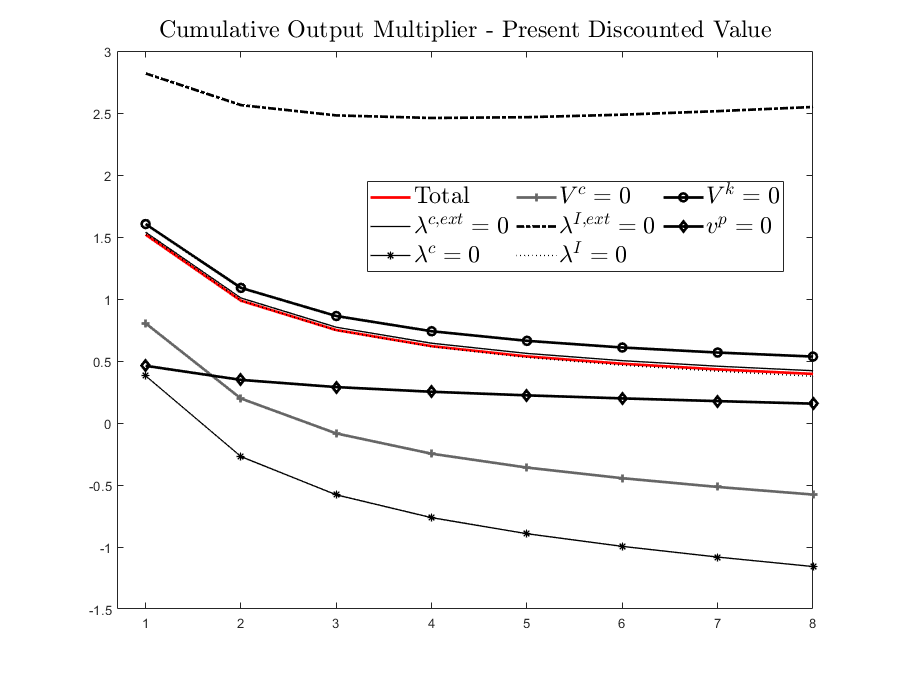} 	\vspace{-0.2 in} 	 		\par	\end{centering} \vspace{-0.15 in} 
	\protect\caption{Counterfactuals at Posterior Mode for $\Theta$}
	\label{out_irf}
\end{figure}

\section{Conclusion}
This paper has shown that a heterogeneous agent model can be approximated around the complete markets allocation and this induces an equilibrium law of motion where higher order moments can be explicitly accounted for. The approximation used lends itself to conventional likelihood based methods, useful for linking the model to a variety of aggregated micro and macro data over the business cycle.  The method was employed to study extensively the effects of financial frictions in the Spanish economy and how different margins depend on aggregate shocks. The semi-structural model was informed using aggregated quantitative and qualitative data which permit the identification of this dependence. Empirical results show that heterogeneity, whether it is cross sectional dispersion or differential response between constrained and unconstrained agents are important empirical determinants of the fiscal multiplier.
 \begin{spacing}{1.2}
\bibliographystyle{econometrica}
\bibliography{biblio_thesis}\end{spacing}
\newpage
\section{Appendix (For Online Publication)}

\subsection{Derivations with external Habits in Consumption}
With external habit formation, the utility of the household becomes as follows: \[u(c_{i,t},l_{i,t},C_{t}):= \left(\frac{\left(c_{i,t}-hC_{t-1}\right)^{1-\omega }-1}{1-\omega }- \frac{l_{i,t}^{1+\eta}}{1+\eta}  \right)  \] and the corresponding Euler equations are: 
\begin{eqnarray*}
	\eta_{i,t}\left(c_{i,t}-hC_{t-1}\right)^{-\omega}&=&\beta\mathbb{E}_{t}\left(c_{i,t+1}-hC_{t}\right)^{-\omega}R^{k}_{i,t+1} + v_{i,t} +\varphi_{i,t}\mu_{i,t}\\
	\left(c_{i,t}-hC_{t-1}\right)^{-\omega}&=&\beta\mathbb{E}_{t}\left(c_{i,t+1}-hC_{t}\right)^{-\omega}\Pi^{-1}_{t+1}R_{t+1} + \mu_{i,t}
\end{eqnarray*}

Expanding individual consumption around aggregate consumption: \begin{eqnarray*}
	&&\int\left(c_{i,t+1}-hC_{t}\right)^{-\omega}d\Lambda_{t}(i)\\
	&\approxeq& \left(C_{t+1}-hC_{t}\right)^{-\omega}\left(1+\frac{\omega(\omega+1)}{2}(C_{t}-hC_{t-1})^{-2}Var_{t}(c_{i,t})\right)\\
	&:=& \left(C_{t+1}-hC_{t}\right)^{-\omega}\Xi_{t}(h)
\end{eqnarray*}
In particular, the bond Euler equation becomes:
\begin{eqnarray*}
	\left(C_{t}-hC_{t-1}\right)^{-\omega}\Xi_{t}(h)&=&\beta\mathbb{E}_{t}\left(C_{t+1}-hC_{t}\right)^{-\omega}\Xi_{t+1}(h)\Pi^{-1}_{t+1}R_{t+1} + \int\mu_{i,t}d\Lambda_{t}(i)
\end{eqnarray*}
Correspondingly, the aggregate intratemporal condition becomes: \begin{eqnarray*}\int l_{i,t} d\Lambda_{t}(i)&=&w_{t}^{\frac{1}{\eta}}\left(c_{i,t}-hC_{t-1}\right)^{-\frac{\omega}{\eta}}\\
	&=&	w_{t}^{\frac{1}{\eta}}(C_{t}-hC_{t-1})^{-\frac{\omega}{\eta}}\left(1+\frac{\omega(\omega+\eta)}{2\eta^{2}}(C_{t}-hC_{t-1})^{-2}Var_{t}(c_{i,t})\right)\\
	&=& w_{t}^{\frac{1}{\eta}}(C_{t}-hC_{t-1})^{-\frac{\omega}{\eta}}\Xi_{t}^{lab}\end{eqnarray*}
The corresponding log-linearized versions for heterogeneity are:  
\[\tilde{\Xi}_{t}=\frac{\omega(1+\omega)Var_{c,ss}}{2(1-h)^{2}+\omega(1+\omega)Var_{c,ss}}\left(\widetilde{Var}_{t}\left(\frac{c_{i,t}}{C_{t}}\right)-\frac{2h}{1-h}(\tilde{C}_{t}-\tilde{C}_{t-1})\right)\] and \[\tilde{\Xi}^{lab}_{t}=\frac{\omega(\eta+\omega)Var_{c,ss}}{2\eta^{2}(1-h)^{2}+\omega(\eta+\omega)Var_{c,ss}}\left(\widetilde{Var}_{t}\left(\frac{c_{i,t}}{C_{t}}\right)-\frac{2h}{1-h}(\tilde{C}_{t}-\tilde{C}_{t-1})\right)\]
\subsection{Priors} 
\begin{flushleft} 	
	\begin{table}[H] 
		\begin{centering}
			\caption{Priors for $\Theta_{FI}$ and Calibration}\vspace{0.1 in}
			\label{table:ParamPriors}
			\begin{tabular}{c|c|c|c|}
				\hline
				$\Theta$& Prior& 	$\Theta$ & Prior\\ \hline 
				$\omega$ &  $\Gamma(3,1)$ &  $\eta$ &$ \Gamma(3,1)$ \\
				$ \beta$& $U(0.98,1)$ &	$h$ & $Beta(1.2,1.2)$\\  
				$\rho_{z} $&$U(-1,1)$&  	$\phi_{d}$ &$U(-60,60)$ \\				
				$\rho_{g}$ &$U(-1,1)$&		    $\phi_{f} $&$U(-60,60)$ \\
				$\rho_{nfl}$&$U(-1,1)$& 	$\sigma_{\epsilon_{Tr}}$ & $InvGamm(0,5)$\\   
				$\rho_{\tau}$ &$U(-1,1)$& 		$\rho_{mpk}$ & $U(-1,1)$	 \\
				$c_{gz}$& $N(0,5)$&    $\rho_{c}$ & $U(-1,1)$ \\  
			    $\rho_{v^{p},(g,m)}$& $N(0,5)$&   $ \rho_{v^{p},z}$& $N(0,5)$ \\  
				$\sigma_z$ & $InvGamm(0,5)$& 	$\rho_{vr}$ & $U(-60,60)$  \\ 
				$\sigma_g$& $InvGamm(0,5)$&  	 $\sigma_{V^{k}}$&$InvGamm(0,5)$\\ 	
				$\sigma_{nfl}$& $InvGamm(0,5)$&      	$pd_{ss}$& $N(0,5)$\\	
				$\sigma_{\tau}$ & $InvGamm(0,5)$&     $\rho_v^{p}$ &$U(-1,1)$ \\ 
				$\lambda \textit{ and } \lambda^{ext}$ & $N(0,5)$ & $\sigma_{v^{p}}$ & $InvGamm(0,5)$ \\ 	 
				$V^{c}_{ss}$ & $InvGamm(0,5)$ & 		$vc$ &  $N(0,5)$\\  	
				$\alpha$ & $0.36$ & 		$\delta$ &  $0.0175$\\			
							\hline
			\end{tabular}
			\par\end{centering}  
		
	\end{table}
\end{flushleft}
\vspace{-0.4 in}
\newpage
\subsection{Survey Data and Transformations}
The question asked to industrial firms in the BCS survey is under the heading "Factors affecting investment in $t$ and $t+1$" and is about the availability of financial resources and their (opportunity) cost, worded as follows : \textit{"This heading covers the availability of resources for investment (and their cost) together with the return  on  investment  and  the  lack  of  opportunities  for  the  company  to  use  its resources  more profitably than by investment (notably by purely financial operations)"}. This akin to an investment efficiency shock in the reference model which makes an investment costly e.g. $\eta_{i,t}>1$. 

Answers to this question are in the form of a balanced statistic over 5 categorical answers, from "very limiting" to "very stimulating". We thus need to transform the series to compute the actual proportion of firms that face this inefficiency. Each answer is chosen by a proportion $p_i$ of firms and the balance statistic $B = p_5+0.5p_4-p_2+0.5p_1$ corresponds to judging whether there is more mass above or below the median $x$, $med(x)$ and truncating the distribution on both ends by the same proportion to avoid extremes. Taking the truncated distribution as the true distribution, then the balance statistic is $B\equiv \mathbb{P}(x>med(x))+\mathbb{P}(x<med(x))$ which implies that $\mathbb{P}(x>med(x))=\frac{1+B}{2}$. This measure is at annual frequency and is treated as a noisy measure of the true proportion, where the variance of the measurement error is calibrated at $1\%$ of the variance.
\subsection{Consumption Dispersion: SHF versus HBS}
Two measures of durable and non durable consumption dispersion are used using data from two different surveys. More specifically, for the SHF, we compute the variance of annual nominal consumption, using survey weights, computed as follows: \small \begin{eqnarray*}
	c^{nom,annual}_{i,t} &=& 12*(\textit{Imputed Monthly Rent} [code: 2.31]\\
	&& + \textit{Car Purchases}/12 [code: 2.74] + \textit{House Durables}/12 [code:p2.70]\\
	&& +\textit{monthly non durable consumption} [code:p9.1])\end{eqnarray*} \normalsize To convert it to the real consumption ratio, I divide by aggregate consumption. 
For the Household Budget Survey, I compute the variance of total consumption expenditure (\textit{code: GASTOT}, for those who do not report zero total income), which is itself a weighted measure of consumption expenditure for the \textit{representative household} of a particular classification. Notice also the survey switched from the COICOP to the ECOICOP classification in 2016.   
 \vspace{-0.1 in}
\subsection{Tax Rates}
The raw measures of tax rates broadly co-move before 2009, while after 2009 it becomes statistically insignificant \small ($Corr \approxeq 0.75,CI_{95\%}= (0.171,0.944)
$) and \small ($Corr \approxeq -0.16, CI_{95\%}=(-0.717,0.522
) $) respectively.\normalsize \vspace{-0.15 in}
\begin{figure}[H]
	\begin{centering}
		\includegraphics[scale=0.33]{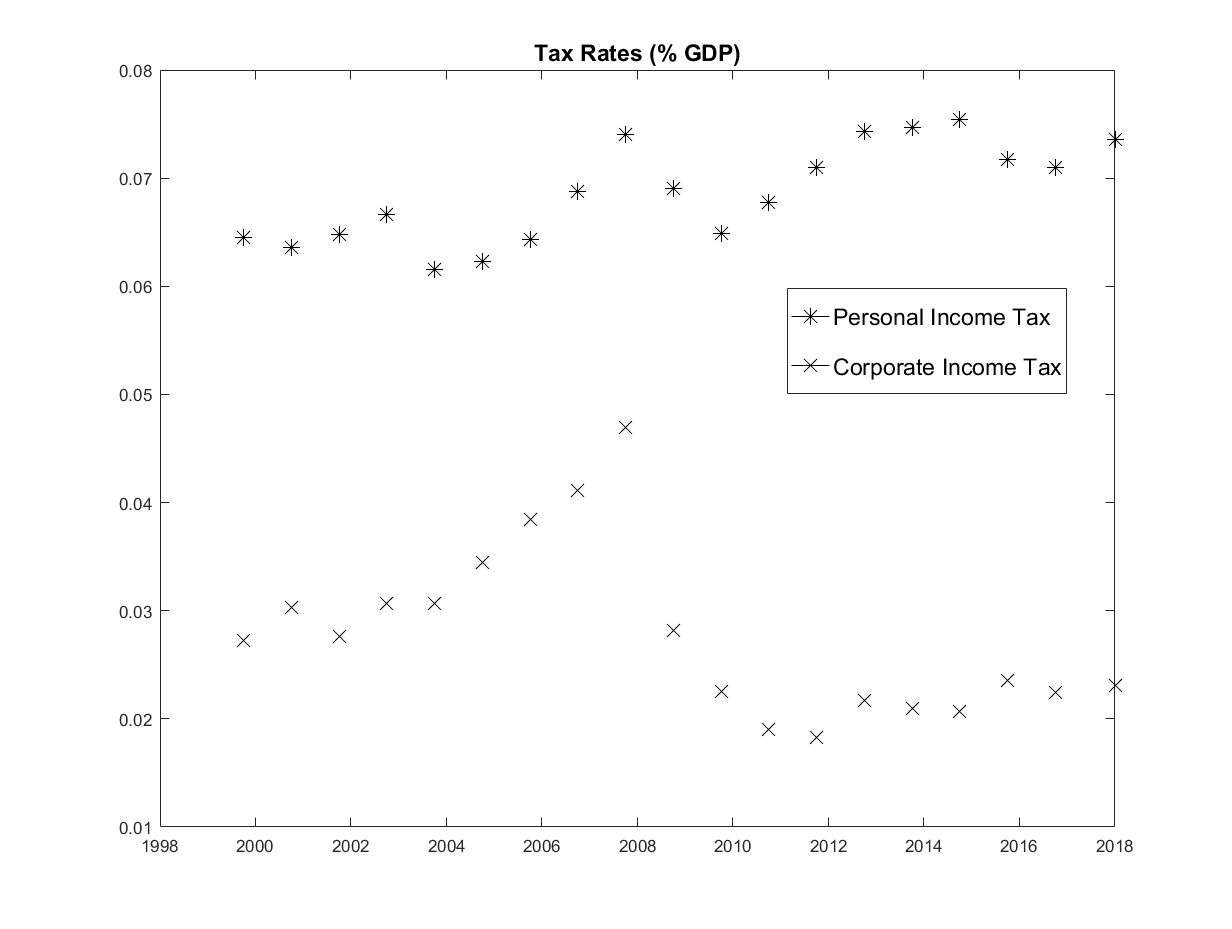}
		\par
	\end{centering}   \vspace{-0.2 in}
	\protect\caption{}
	\label{taxrates}
\end{figure}

 \newpage
\subsection{Proofs}  
\begin{proof}
	\textbf{of Proposition 1}.
	The model with frictions is represented as follows:  \small
	\begin{eqnarray*}
		G(\theta_1,0)X_{t}&=&F(\theta_1,0)\mathbb{E}_{t}(X_{t+1}|X_{t})+L(\theta_1,0)Z_{t}+{\mu _{t}}
	\end{eqnarray*}\normalsize
	Plugging in the candidate distorted decision rule: $X_{t}=X_{t}^{\star}+{\lambda}_{t}$ and using the competitive equilibrium conditions for the frictionless model \vspace{-0.1 in}\[\underset{n_{x}\times n_{x}}{F(\theta_1,0)}\underset{n_{x}\times n_{x}}{P^{\ast }(\theta_1,0)}-\underset{n_{x}\times n_{x}}{G^{\ast }(\theta_1,0)}=0\] \[(\underset{n_{z}\times n_{z}}{R(\theta_1,0)^{T}}\otimes F(\theta_1,0) + I_z\otimes(F(\theta_1,0)P^{\ast }(\theta_1,0)-G^{\ast }(\theta_1,0)))vec(Q(\theta_1,0))=-vec(L(\theta_1,0))\]
	leads to the following condition:\vspace{-0.1 in} \[G(\theta_1,0){\lambda}_{t}=\mathbb{E}_{t}F(\theta_1,0){\lambda}_{t+1} + {\mu _{t}}\]
	Substituting for ${\mu}_{t}=M_{1}X_{t-1}+M_{2}Z_{t}$ and the guess ${\lambda}_{t} = H {\mu}_{t}$ and matching coefficients on $X_{t-1}$ and $Z_{t}$ we arrive at 	\begin{eqnarray*} vec(H) &=& \left(M_{1}'\otimes G(\theta_{1},0) - (M_{1}P^{\star}(\theta_{1},\theta_{2}))'\otimes F(\theta_{1},0)\right)^{-1}vec(M_{1})\\
		vec(M_{2}) &=& \left(I_{n_{z}}\otimes (G(\theta_{1},0)H-I_{n_{x}})-R'\otimes F(\theta_{1},0)H  \right)^{-1}\times\\
		&&\left((M_{1}Q^{\star}(\theta_{1},\theta_{2}))'\otimes F(\theta_{1},0)\right)vec(H)\end{eqnarray*}
\end{proof}

\subsection{\textbf{Experiment 2}: Adverse parameterization $(\omega=1.5$, $\beta = 0.97$)}
In this experiment I lower both the discount factor and the risk aversion coefficient. This generates weaker motives for precautionary savings, and the expectation is that agents will now be more exposed to idiosyncratic risk. The question is therefore whether aggregates are still well approximated empirically despite the fact that the variance of consumption shares cannot be predicted perfectly by aggregate shocks. 

As evident from the figures below, this is the case. Yet this can be explained by "approximate aggregation". Since the distribution does not matter, then it should not matter if the model cannot perfectly track the evolution of the variance using the aggregate state variables. 
\begin{figure}[H]
	\includegraphics[scale=0.42]{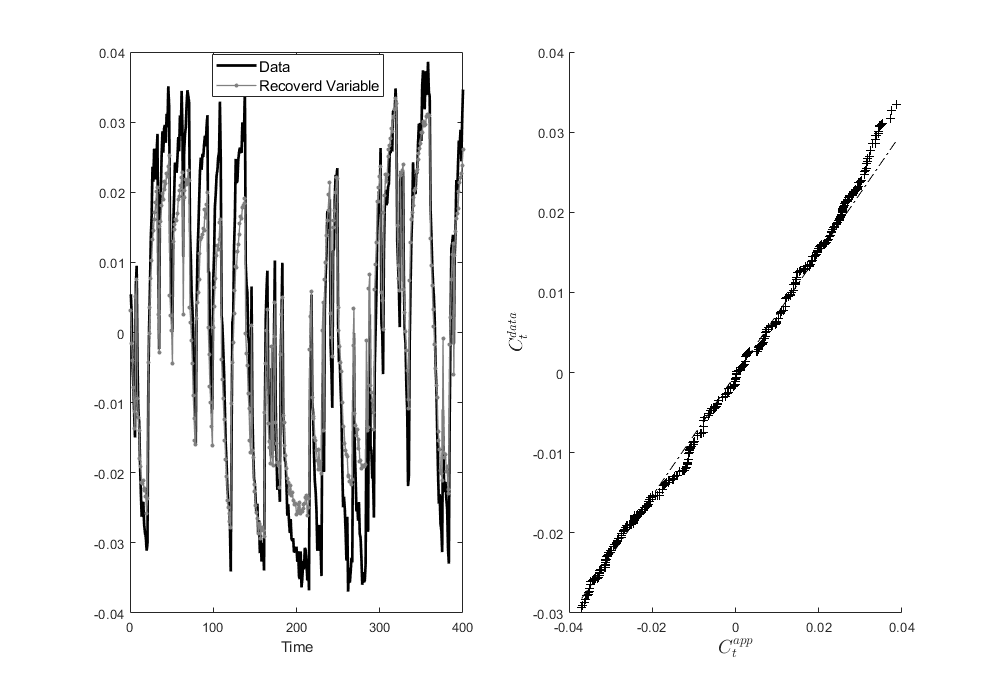}
	\protect\caption{Aggregate Consumption}
\end{figure} \vspace{-0.2 in}

\begin{figure}[H]
	\includegraphics[scale=0.42]{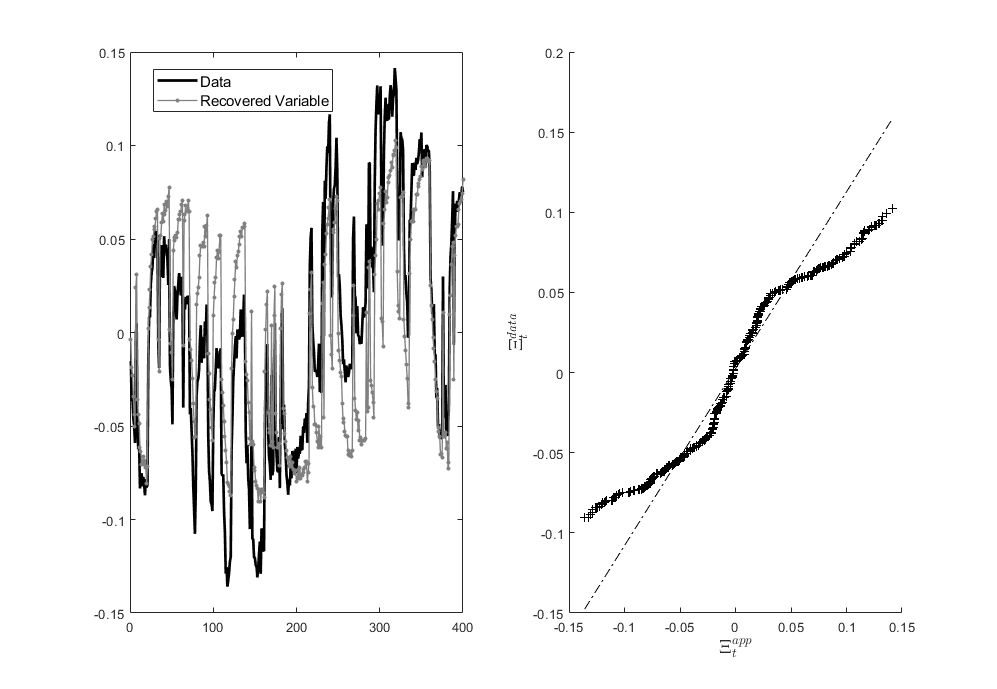}
	\protect\caption{Variance of Consumption Shares}
\end{figure} \vspace{-0.2 in}
\begin{figure}[H]
	\includegraphics[scale=0.42]{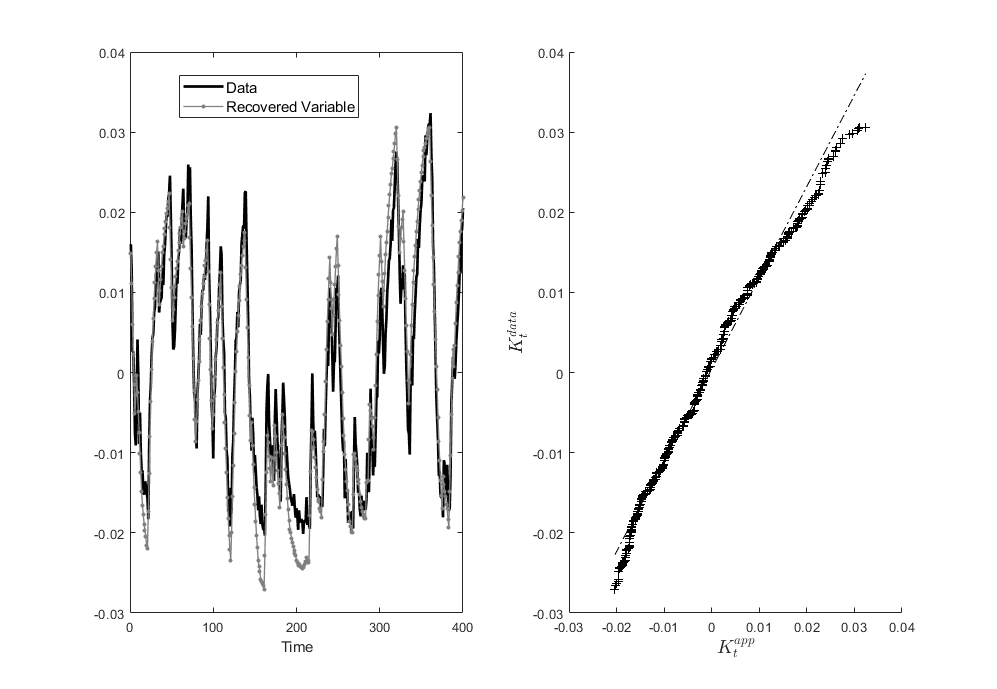}
	\protect\caption{Aggregate Capital Stock}
\end{figure} \vspace{-0.2 in}

%

%
%
%
%

\newpage
%
%
%
%

\subsection{Mixed Frequency Model for Quarterly Proportion of $\tilde{B}^{con}_{t}$}

Using  $\tilde{B}^{con}_{t} = \tilde{p}_{t}+\tilde{\zeta}_{t}$ as a measurement equation, the true quarterly proportion of liquidity constrained consumers is extracted using the following mixed frequency Gaussian linear state space model, where $t=4(j-1)+q$, $t$ is the quarterly observation at year $j$ and $q=\{1,2,3,4\}$ the within year quarter index:\newline
State Equation (ignoring identities): 
\begin{eqnarray*}
	\left( \begin{array}{c}
		\tilde{B}^{con}_{4(j)}\\
		\tilde{\zeta}_{4(j)}\\
	\end{array}\right) &=&  \left(\begin{array}{cc}
		\rho_{p} & 0 \\
		0 & \rho_{\zeta} 
	\end{array}\right)  \left( \begin{array}{c}
	\tilde{B}^{con}_{4(j-1)+3}\\
		\tilde{\zeta}_{4(j-1)+3}\\
	\end{array}\right) + \left(\begin{array}{c}
		\nu_{\tilde{B}^{con},4(j)}\\
		\nu_{\tilde{\zeta},4(j)}
	\end{array}\right)\end{eqnarray*}
Observation equation:
\begin{eqnarray*}
	\left(\begin{array}{c}
		\tilde{p}^{o}_{4(j)}\\
		\tilde{B}^{con,o}_{j}\end{array}\right)&=& \left( \begin{array}{cccccccccccc}
		1 & 0 & 0 &0 & 0 & 0 &0 & 0 &1& 0 &... & 0\\
		\frac{1}{8} &\frac{1}{8} & \frac{1}{8} &\frac{1}{8} & \frac{1}{8} & \frac{1}{8} &\frac{1}{8} &\frac{1}{8}&0&0&...&0 \\
	\end{array} \right)\left( \begin{array}{c}
		\tilde{B}^{con}_{4(j)}\\
		\tilde{B}^{con}_{4(j-1)+3}\\
		\tilde{B}^{con}_{4(j-1)+2}\\
		\tilde{B}^{con}_{4(j-1)+1}\\
		\tilde{B}^{con}_{4(j-1)}\\
		\tilde{B}^{con}_{4(j-2)+3}\\
		\tilde{B}^{con}_{4(j-2)+2}\\
		\tilde{B}^{con}_{4(j-2)+1}\\
		\tilde{\zeta}_{4(j-1)+3}\\
		\tilde{\zeta}_{4(j-1)+2}\\
		\tilde{\zeta}_{4(j-1)+1}\\
		\tilde{\zeta}_{4(j-2)}\\
		\tilde{\zeta}_{4(j-2)+3}\\
		\tilde{\zeta}_{4(j-2)+2}\\
		\tilde{\zeta}_{4(j-2)+1}\\
		\tilde{\zeta}_{4(j-2)}
	\end{array}\right) + v_{4(j)} 
\end{eqnarray*} 
where  $(\nu_{\tilde{B}^{con},t},\nu_{\tilde{\zeta},t})\sim N(0,diag(\Sigma_{\nu}))$ and $v_{t}\sim N(0,diag(\Sigma_{v}))$. The last diagonal component of $\Sigma_{v}$ is calibrated to the standard error from the estimation of the proportion of constrained consumers in the SHF survey. The rest of the components of the diagonal are calibrated to $1\%$ of the variance of the BCS measure ($\tilde{p}_{t}$).

\subsection{Model for Trends} Let $(X^{1,obs}_{t}, X^{2,obs}_{t}, X^{3,obs}_{t})'$ denote the three groups of observables, where the first two groups have a common group specific trend: $\Upsilon_{t} = (\Upsilon^{1}_{t} ,\Upsilon^{2}_{t} )$. The state space formulation is as follows:\newline
State Equation:
\begin{eqnarray*}
		\left(\begin{array}{c}
	     \Upsilon_{t}\\
	     X^{np}_{t}\end{array}\right)	     
	      & = & \left(\begin{array}{cc}
	      	\mathcal{I}&0\\
	      	0&P^{\star}(\theta)
	      	\end{array}\right)\left(\begin{array}{c}
	      	\Upsilon_{t-1}\\
	      	 X^{np}_{t-1}\end{array}\right) +     	
	      		\left(\begin{array}{c}	      	
	      	 \epsilon^{p}_{t}\\
            Q^{\star}(\theta)\epsilon^{np}_{t}\end{array}\right)
	\end{eqnarray*} 
Observation equation:
\begin{eqnarray*}
		X^{obs}_{t} & = & \mathcal{S}	\left(\begin{array}{c}
			\Upsilon_{t}\\
			X^{np}_{t}\end{array}\right)	 +  \epsilon^{me}_{t}
\end{eqnarray*} 
where  $(\epsilon^{p}_{t})\sim N(0,\Sigma_{p})$. The non-zero elements of $\Sigma_{p}$ are those that correspond two groups and $\mathcal{S}=blockdiag(S_{\Upsilon},S_{x})$. $S_{x}$ is a selection matrix. The specificatoin accomodates for trends at quarterly and yearly frequency. 

\begin{landscape}
\small
\begin{eqnarray*}
	\mathcal{I}&=& \left(
\begin{array}{cccc|c c c c}
	 1&0&0& & & & &\\
	 1&0&0& & & & \mathbf{0}_{4}&\\
	 0&1&0&&&&\\
	 0&0&1&&&&\\
	  &  & &  & 1 & 0 & 0 & \\
	   &  \mathbf{0}_{4}&  & & 1 & 0 & 0 &  \\
	  &  & &   & 0 & 1 & 0 &  \\
	  &  & &   & 0 & 0 & 1 &  \\
		 \end{array}
\right)\end{eqnarray*}
\begin{eqnarray*}
	S_{\Upsilon}&=& \left(
	\begin{array}{cccc|c c c c}
		1&0&0& 0& & & &\\
		1&0&0&0 & & & \mathbf{0}_{4}&\\
		1&0&0&0&&&\\
		1&0&0&&&\\
		s_{c}&s_{c}(1+2\rho_{c})  &s_{c}(1+2\rho_{c}(1+\rho_{c})) & s_{c}(1+2\rho_{c}(1+\rho_{c}+\rho^{2}_{c})) & 0 & 0 & 0 & 0\\
		s_{c}&s_{c}(1+2\rho_{c})  &s_{c}(1+2\rho_{c}(1+\rho_{c})) & s_{c}(1+2\rho_{c}(1+\rho_{c}+\rho^{2}_{c})) & 0 & 0 & 0 & 0\\
		0& 0 &0 &0   & s_{mpk} & 0 & 0 & 0 \\
		0& 0 &0 &0   & 1 & 1+2\rho_{mpk} & 1+2\rho_{mpk}(1+\rho_{mpk})&  1+2\rho_{mpk}(1+\rho_{mpk}+\rho^{2}_{mpk}) \\
	\end{array}
	\right)\end{eqnarray*}

\normalsize

\end{landscape}
\section{Estimated Trends}
Below I plot the estimated trends in consumption level and dispersion, government spending, wages and investment as well as the common trend in the ex ante real rate and the dispersion in the marginal revenue product. 
 \begin{figure}[H]
	\begin{centering}
		\includegraphics[scale=0.42]{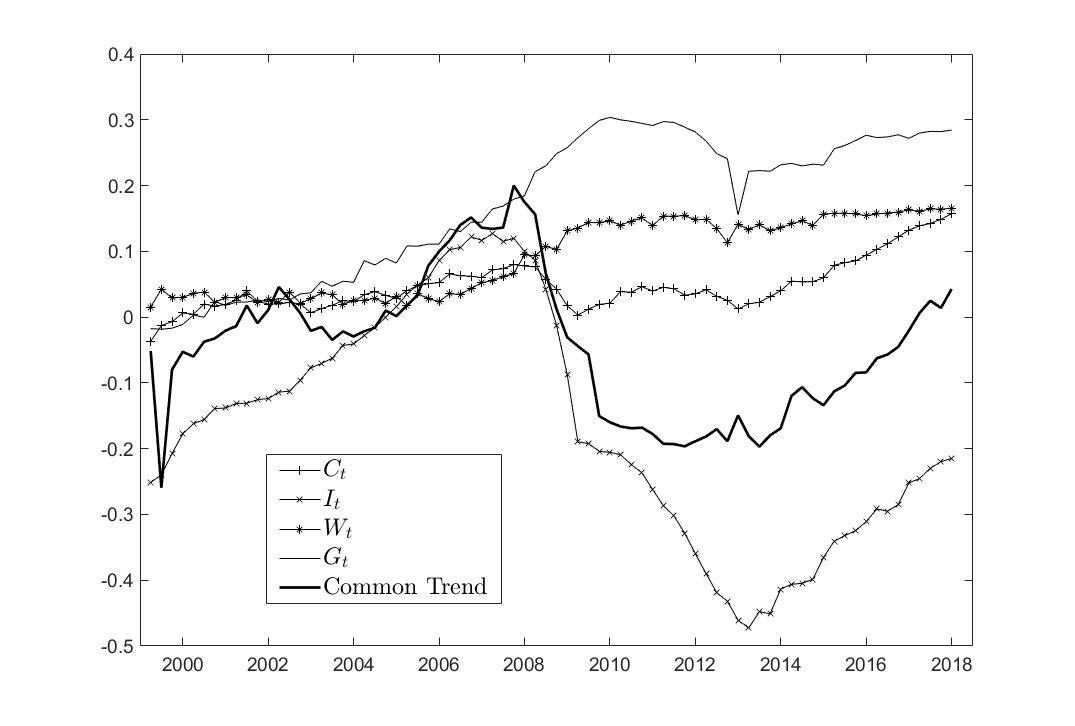}
		\includegraphics[scale=0.42]{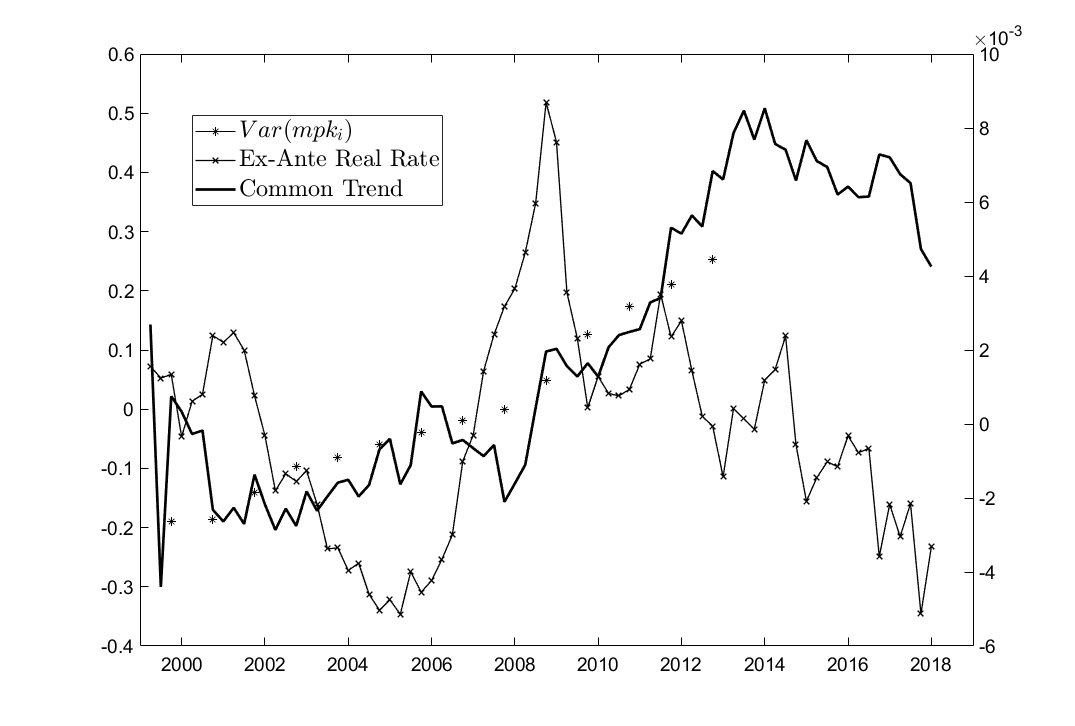}
		\par
	\end{centering}   \vspace{-0.2 in}
	\protect\caption{}
	\label{Trends}
\end{figure}

\begin{landscape}
\section{Assessing Fit}$ $
\vspace{-0.4 in}
\begin{figure}[H]
	\begin{centering}
		\includegraphics[scale=0.65]{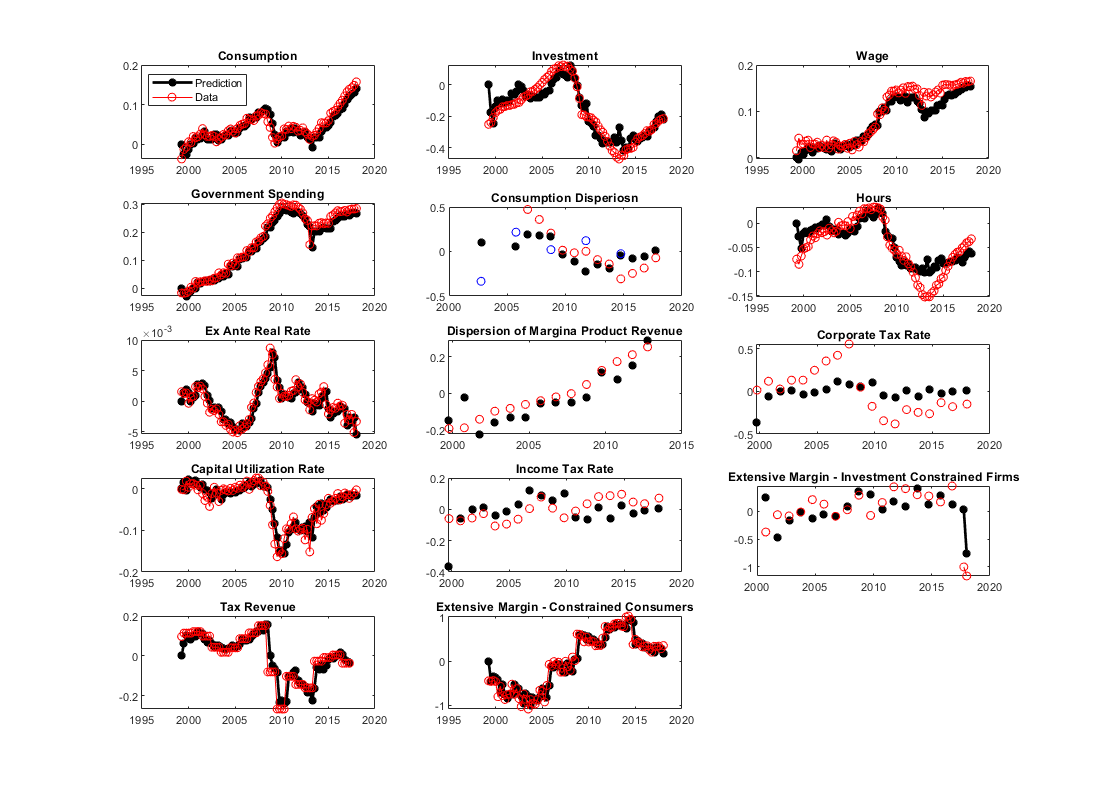}
		\par
	\end{centering}   \vspace{-0.4 in}
	\protect\caption{Data against the model based predictions using the Kalman filter.}
	\label{predict}
\end{figure}
\end{landscape}

\section{Consumption and Investment Fiscal Multipliers}$ $
\begin{figure}[H]
	\begin{centering}
		\includegraphics[scale=0.45]{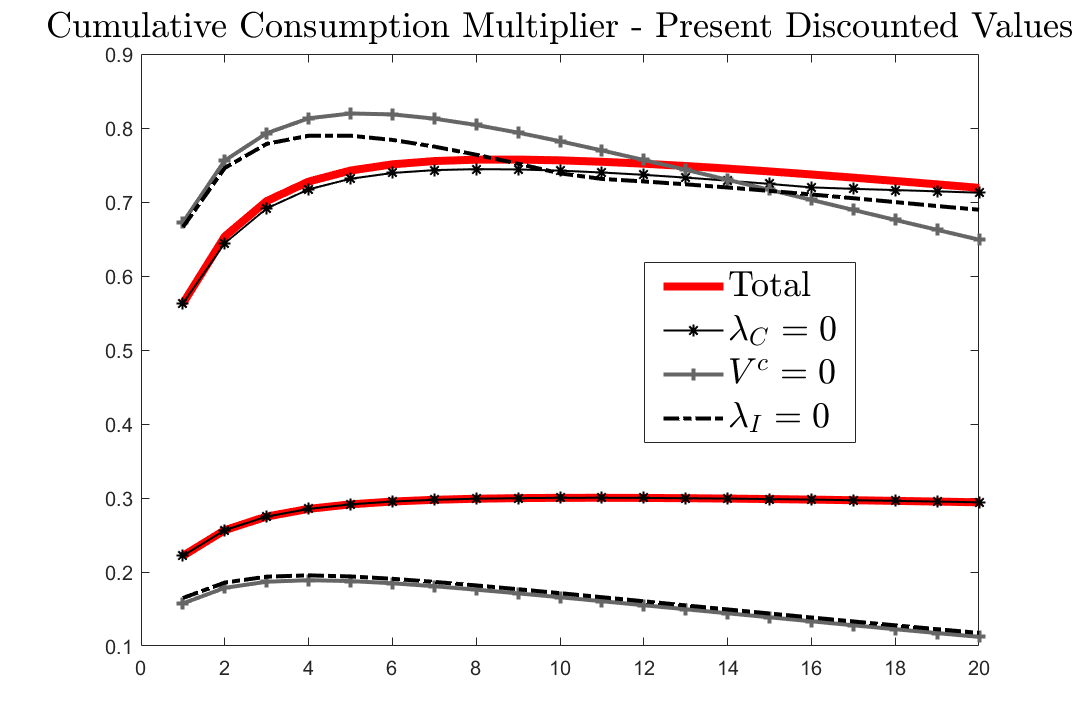}
			\includegraphics[scale=0.45]{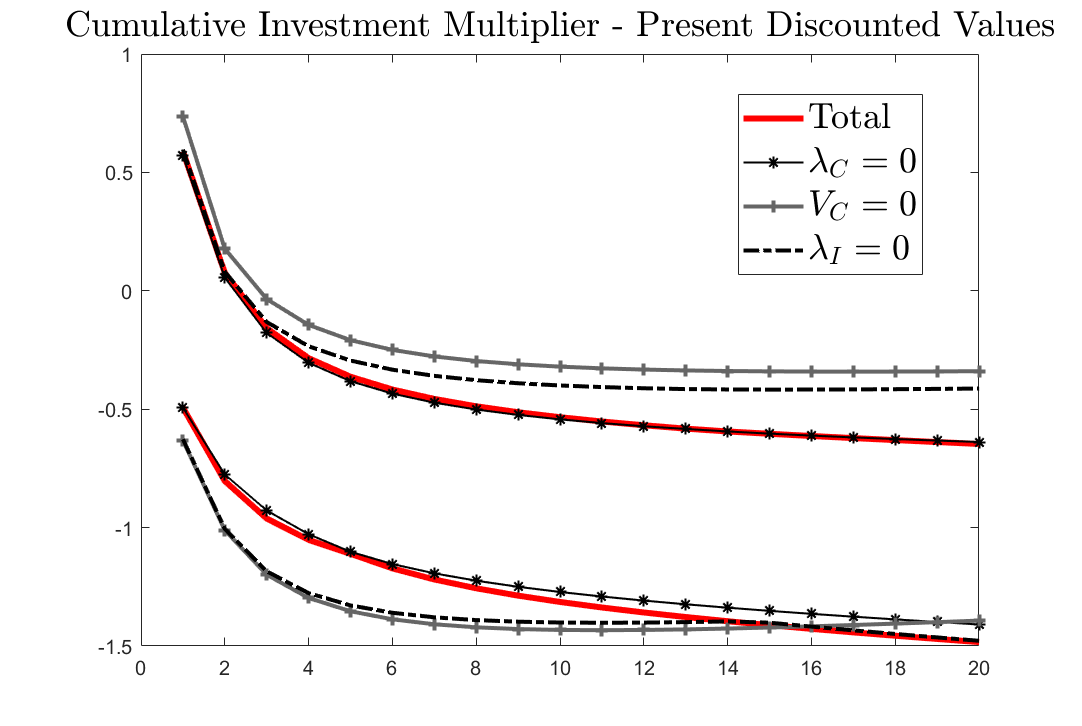}   	\vspace{-0.15 in} 	 		\par	\end{centering} \vspace{-0.15 in} 
	\protect\caption{$90\%$ Confidence Sets}
	\label{cons_mult}
\end{figure}

\begin{figure}[H]
	\begin{centering}
			\includegraphics[scale=0.55]{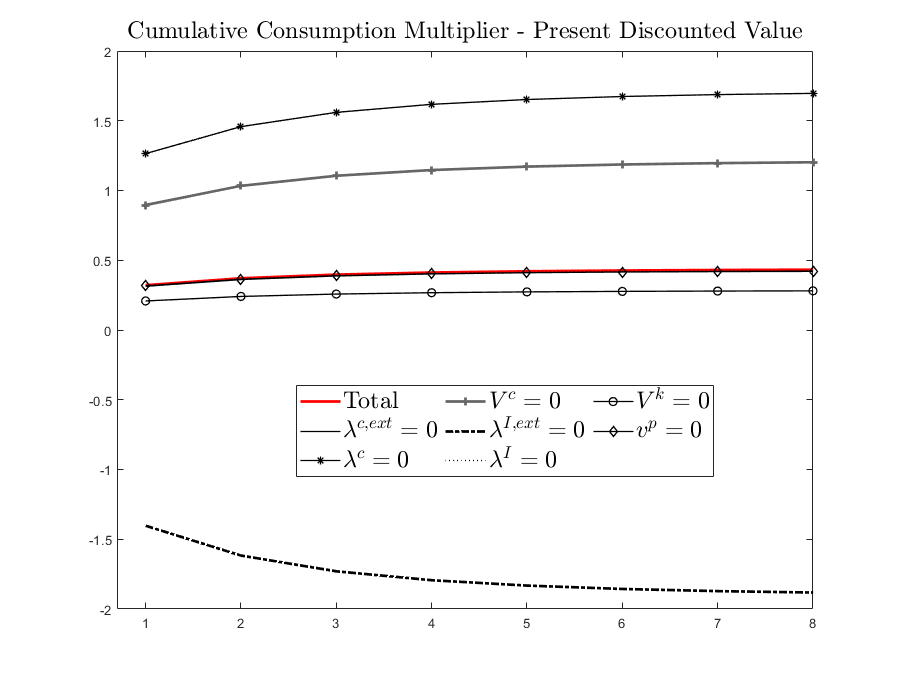}   
				\includegraphics[scale=0.55]{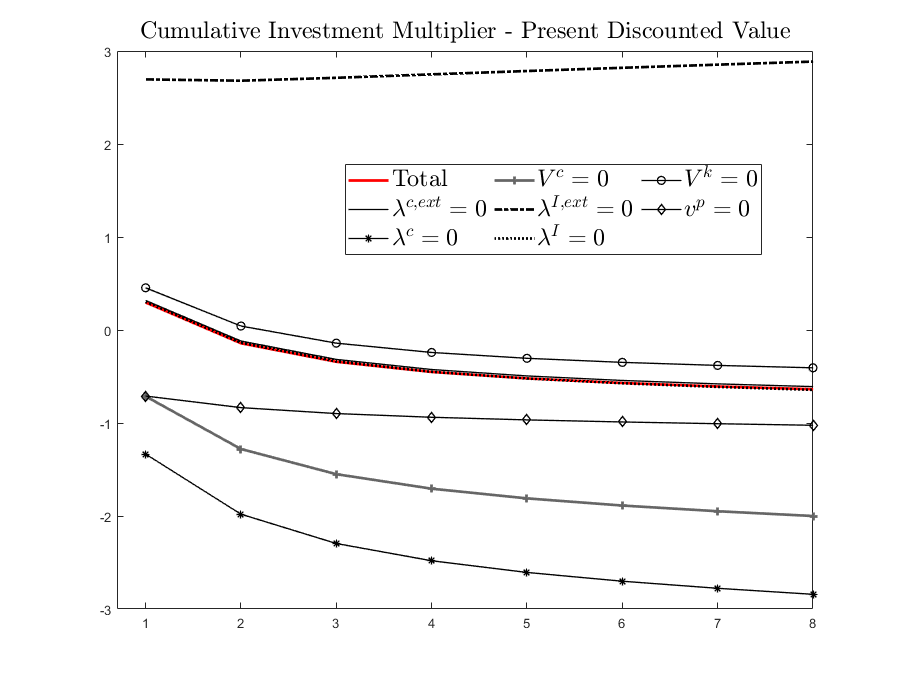}   
				
					 		\par	\end{centering} \vspace{-0.15 in} 
	\protect\caption{Counterfactuals at Posterior Mode for $\Theta$}
	\label{cons_irf}
\end{figure}
\section{Augmented Model with Capital Euler Equation}
Augmenting the model with the Capital Euler Equation is not expected to offer much in terms of identification of the existing parameters in the empirical model of section 7, unless one is directly interested in the determination on the distortions to the value of capital. Recall that the  individual first order condition (without habits) is as follows: 
\begin{eqnarray*}
	\eta_{i,t}c^{-\omega}_{i,t}&=&\beta\mathbb{E}_{t}c^{-\omega}_{i,t+1}((1-\delta)\eta_{i,t+1} + (1-\tau_{t+1}) \frac{\psi}{2}u^{2}_{i,t+1}) + v_{i,t} +\varphi_{i,t}\mu_{i,t}
\end{eqnarray*} 
A different way to determine the aggregate implications of inefficiency in capital accumulation is to re-write the Euler equation in terms of the price of capital.
Dividing the first equation by marginal utility, applying the same type of second order approximation and aggregating using the joint $\Lambda_{t,t+1}$ distribution and size weights $\frac{k_{i,t}}{K_{t}}$, we get that $
 Q_{t} =\beta\mathbb{E}_{t}\left(\frac{C_{t+1}}{C_{t}}\right)^{-\omega}\frac{\Xi_{t+1}}{\Xi_{t}}R^{k}_{t+1} + \mu^{Q}_{t}$ where \small \[Q_{t}:=\int \eta_{i,t}\frac{k_{i,t}}{K_{t}}d \Lambda_{i}(t),\mu^{Q}_{t}:=\int c^{\omega}_{i,t}\varphi_{i,t}\mu_{i,t}\frac{k_{i,t}}{K_{t}}d \Lambda_{i}(t)\] and
	\[R^{k}_{t}= \int R^{k}_{i,t}\frac{k_{i,t}}{K_{t}} d\Lambda_{i}(t)= (1-\delta)Q_{t} + (1-\tau_{t}) \left(\frac{\psi}{2}\alpha \frac{Y_{t}}{K_{t}}\frac{\nu^{p}_{t}-1}{\nu^{p}_{t}}\right)\]  \normalsize
	The distortion $\mu^{Q}_{t}$ affects the aggregate value of the capital stock, which will then have to be linked to all shocks as in the rest of the cases.	Moreover, to see that the distortion $\lambda_{I,t}$ can be linked to $Q_{t}$, notice that:	\small
	\begin{eqnarray*}
\lambda_{I,t} &=&  \int \iota_{i,t}\frac{\eta_{i,t}-1}{\eta_{i,t}}d\Lambda_{t}(i)\approx \frac{I_{t}}{Q_{t}}(Q_{t}-1 -V^{Q}_{t})
	\end{eqnarray*} \normalsize
where $V^{Q}_{t}$ is the cross sectional dispersion of $\eta_{i,t}$. Assuming that the ex-ante real rate is actually known at time $t$, we can also see that $Q_{t}=\mathbb{E}_{t}\frac{R^{k}_{t+1}}{R_{t+1}}+\mu^{Q}_{t}$.
 A positive discrepancy between the return to capital and the real interest rate tomorrow as well as borrowing constraints imply that $Q_{t}$ can be well above one. Thus, the aggregate shocks that affect the after tax revenue product of capital and the distortion $\mu^{Q}$ will be the key shocks that  distort aggregate investment downwards.

\end{document}